\documentclass[prx,aps,onecolumn,amsfonts,showpacs,superscriptaddress,preprint,eqsecnum]
{revtex4-1}

\usepackage{amsmath,amssymb,bm} 
\usepackage{graphicx}
\usepackage{color}
\usepackage{subfigure} 
\usepackage{braket}
\usepackage[colorlinks=true]{hyperref}  
\hypersetup{
    bookmarks=true,         
    unicode=false,          
    pdftoolbar=true,        
    pdfmenubar=true,        
    pdffitwindow=false,     
    pdfstartview={FitH},    
    pdftitle={Quantum Quench of the Sachdev-Ye-Kitaev Model},    
    pdfauthor={Andreas Eberlein, Valentin Kasper, Subir Sachdev, Julia Steinberg},     
    pdfsubject={},   
    pdfcreator={},   
    pdfproducer={}, 
    pdfkeywords={} {} {}, 
    pdfnewwindow=true,      
    colorlinks=true,       
    linkcolor=magenta, 
    citecolor=blue,        
    filecolor=magenta,      
    urlcolor=blue           
} 
\usepackage{amsfonts}
\usepackage{upgreek}
\usepackage{slashed}
\usepackage{latexsym}
\usepackage{braket}

\usepackage[utf8]{inputenc}

\usepackage{todonotes}

\newcommand{\beq}{\begin{equation}}
\newcommand{\eeq}{\end{equation}}
\def\bea{\begin{eqnarray}}
\def\eea{\end{eqnarray}}

\newcommand{\nn}{\nonumber \\}

\newcommand{\ie}{\textit{i.\,e.}}

\begin{document}

\title{Quantum quench of the Sachdev-Ye-Kitaev Model}

\author{Andreas Eberlein}
\affiliation{Department of Physics, Harvard University, Cambridge, 
Massachusetts 02138, USA}

\author{Valentin Kasper}
\affiliation{Department of Physics, Harvard University, Cambridge, 
Massachusetts 02138, USA}

\author{Subir Sachdev}
\affiliation{Department of Physics, Harvard University, Cambridge, 
Massachusetts 02138, USA}
\affiliation{Perimeter Institute for Theoretical Physics, Waterloo, Ontario, Canada N2L 2Y5}

\author{Julia Steinberg}
\affiliation{Department of Physics, Harvard University, Cambridge, 
Massachusetts 02138, USA}


\date{\today}

\begin{abstract}%
We describe the non-equilibrium quench dynamics of the Sachdev-Ye-Kitaev models of fermions with random all-to-all
interactions. These provide tractable models of the dynamics of quantum systems without quasiparticle excitations. The Kadanoff-Baym equations show that the final state is thermal, and their numerical analysis is consistent
with a thermalization rate proportional to the absolute
temperature of the final state. We also obtain an
exact analytic solution of the quench dynamics
in the large $q$ limit of a model with $q$ fermion interactions: in this limit, the thermalization of the fermion Green's function is instantaneous.
\end{abstract}


\maketitle

\section{Introduction}
\label{sec:intro}

The non-equilibrium dynamics of strongly interacting quantum many-particle systems
have been the focus of much theoretical work \cite{KamenevBook}. These are usually studied
within the Schwinger-Keldysh formalism, which can describe evolution from a generic initial
state to a final state which reaches thermal equilibrium at long times. The thermodynamic
parameters of the final state ({\em e.g.\/} temperature) 
are determined by the values of the  conjugate conserved quantities ({\em e.g.\/} energy). 
The Kadanoff-Baym equations obtained from this formalism describe the manner and rate by which this final thermal state is reached.

The Kadanoff-Baym equations are usually too difficult to solve in their full generality.
Frequently, a quasiparticle structure has been imposed on the spectral functions, so that the Kadanoff-Baym
equations reduce to a quantum Boltzmann equation for the quasiparticle distribution functions. 
Clearly such an approach cannot be employed for final states of Hamiltonians which describe
critical quantum matter {\em without\/} quasiparticle excitations. 
The most common approach is then to employ an expansion away from a regime where quasiparticles 
exist, using a small parameter such as the deviation of dimensionality from the critical dimension,
or the inverse of the number of field components: the analysis is then still carried out using
quasiparticle distribution functions \cite{DS97,Sachdev97}.

In this paper, we will examine the non-equilibrium dynamics towards a final state without quasiparticles.
We will {\em not\/} employ a quasiparticle decomposition, and instead provide solutions of the full Kadanoff-Baym
equations.
We can obtain non-equilibrium solutions for the Sachdev-Ye-Kitaev (SYK) 
models \cite{sy,kitaev2015talk,Maldacena2016} with all-to-all and random 
interactions between $q$ Majorana fermions on $N$ sites. These models are solvable realizations of quantum matter
without quasiparticles in equilibrium, and here we shall extend their study to non-equilibrium dynamics.
We shall present numerical solutions of the Kadanoff-Baym equations for the fermion Green's function at $q=4$, and in Section~\ref{sec:largeq}, an exact analytic
solutions in the limit of large $q$ (no quasiparticles are present in this limit). The large $q$ solution relies on a remarkable exact SL(2,C)
invariance, and has connections to quantum gravity on AdS$_2$ with the Schwarzian effective action \cite{SS10,SS10b,AAJP15,kitaev2015talk,Maldacena2016,KJ16,HV16} for the equilibrium dynamics, as we will note in Section~\ref{sec:conc}.

The equilibrium SYK Hamiltonian we shall study is (using conventions from 
Ref.~\onlinecite{Maldacena2016})
	\begin{equation}
H=(i)^{\frac{q}{2}}\sum_{1\leq i_{1}<i_{2}<...<i_{q}\leq N} 
j_{i_{1}i_{2}...i_{q}}\psi_{i_{1}}\psi_{i_{2}}...\psi_{i_{q}} 
\label{eq:generalq}
\end{equation}
where $\psi_i$ are Majorana fermions on sites $i = 1 \ldots N$ obeying
\beq
\{ \psi_i , \psi_j \} = \delta_{ij} \,,
\eeq
and $j_{i_{1}i_{2}...i_{q}}$ are Gaussian random variables with zero mean and 
variance
\begin{equation}
\langle j^{2}_{i_{1}...i_{q}}\rangle=\frac{J^{2}(q-1)!}{N^{q-1}}.
\label{eq:variance}
\end{equation}
Our numerical analysis of the SYK model will be carried out on the case with   
time-dependent $q=2$ and 
$q=4$ terms:
\begin{equation}
	H(t) = i \sum_{i < j} j_{2,ij} \, f(t) \, \psi_i \psi_j - \sum_{i < j < k < l} 
j_{4,ijkl} \, g(t) \, \psi_i \psi_j \psi_k \psi_l,
\label{eq:Hamiltonian}
\end{equation}
where $j_{2,ij}$ and $j_{4,ijkl}$ are random variables as described above, and 
$f(t)$ and $g(t)$ can be arbitrary functions of time. We will focus
on a particular quench protocol in which
\bea
&& \mbox{for $t < 0$, $f(t) = 1$ and $g(t) = 1$}\,;\nn	
&& \mbox{for $t > 0$, $f(t) = 0$ and $g(t) = 1$}\,.
\eea
For $t<0$, we have a thermal initial state with both $q=2$ and $q=4$ interactions present. In this case the $q=2$ free fermion terms dominate at low energies, and hence we have Fermi liquid
behavior at low temperatures.
For $t>0$, we have only the $q=4$ Hamiltonian as in Eq.~(\ref{eq:generalq}), which describes a 
non-Fermi liquid even at the lowest energies \cite{sy}. Just after the quench at $t=0$, the system
is in a non-equilibrium state. We show from the Kadanoff-Baym equations, and verify by our numerical
analysis, that the $t \rightarrow \infty$ state is indeed a thermal non-Fermi liquid state at $q=4$, which is at an inverse
temperature $\beta_f = 1/T_f$. The value of $\beta_f$ 
is such that the total energy of the system remains the same after the quench at $t=0^+$.

We note in passing that an alternative quench protocol in which $f(t) = 0$ for all $t$, while $g(t)$ is
time dependent (or the complementary case in which $g(t)=0$ for all $t$, while $f(t)$ is time-dependent), does not yield any non-trivial dynamics (this was pointed out to us by J. Maldacena). This case corresponds to an overall rescaling of
the Hamiltonian, which does not change any of the wavefunctions of the eigenstates. So an initial state in thermal equilibrium will evolve simply via a time reparameterization which
is given in Eq.~(\ref{e1a}).

We will describe the time evolution of the system for $t \geq 0$ by computing a number of two-point
fermion Green's functions $G(t_1, t_2)$. As the system approaches thermal equilibrium, it is useful to characterize
this correlator by 
\beq
\mathcal{T} = (t_1 + t_2)/2 \quad, \quad t = t_1 - t_2 \,. \label{eq:wigner}
\eeq
Then we have
\beq
\lim_{\mathcal{T} \rightarrow \infty} G(t_1, t_2) = G_{\beta_f} (t)
\eeq
where $G_{\beta_f}$ is the correlator in equilibrium at an inverse temperature $\beta_f$. 
For large $\mathcal{T}$ we will characterize $G(t_1, t_2)$ by an effective temperature $\beta_{\rm eff} (\mathcal{T})$ (see below). From our numerical analysis we characterize the late time approach
to equilibrium by
\beq
\beta_\text{eff} (\mathcal{T})  = \beta_f + \alpha \exp \left( - \Gamma \mathcal{T} \right)\,. \label{betaGamma}
\eeq
This defines a thermalization rate, $\Gamma$. Our complete numerical results for $\Gamma$ appear in Fig.~\ref{fig:Gamma1},
and 
they are reasonably fit to the behavior 
\beq
\Gamma = C/\beta_f \quad, \quad \beta_f J_4 \gg 1\,, \label{i6}
\eeq
where $C$ is a numerical constant independent of the initial state. Thus, at low final temperatures, the thermalization rate of the non-Fermi liquid state
appears proportional to temperature, as is expected for systems without quasiparticle excitations \cite{ssbook}.

\subsection{Large $q$ limit}
Further, we will also consider a model in which the $t<0$ Hamiltonian has a $q$-fermion interaction $\mathcal{J}$,
and a $pq$ fermion interaction $\mathcal{J}_p$. At $t>0$ we then quench to only a $q$ fermion
interaction $\mathcal{J}$. We will show that the non-equilibrium dynamics of this model are exactly
solvable in the limit of large $q$ taken at fixed $p$. 

The large $q$ Green's function can be written as \cite{Maldacena2016}
\beq
G^> (t_1, t_2) = -i \left\langle \psi (t_1) \psi (t_2) \right\rangle = - \frac{i}{2} \left[ 
1 + \frac{1}{q} g(t_1, t_2) + \ldots \right] \label{e2}
\eeq
where 
\beq
g(t, t) = 0 \,. \label{e2a}
\eeq
For $t_1 >0$ or $t_2 > 0$, we find that $g(t_1, t_2)$ obeys the two-dimensional Liouville equation
\beq
\frac{\partial^2}{\partial t_1 \partial t_2} g(t_1, t_2) = 2 \mathcal{J}^2 \, e^{g (t_1, t_2)} \,,
\label{liouville}
\eeq
where $\mathcal{J}$ is proportional to the final interaction strength (see Eq.~(\ref{e4})).
The most general solution of this equation can be written as \cite{CROWDY}
\beq
g(t_1, t_2)  = \ln \left[ \frac{-h_1' (t_1) h_2' (t_2)}{\mathcal{J}^2 (h_{1} (t_1) - h_{2} (t_2))^2} \right] \, , \label{i3}
\eeq
where $h_1 (t_1)$ and $h_2 (t_2)$ are arbitrary functions of their arguments. A remarkable and significant
feature of this expression for $g(t_1, t_2)$ is that it is exactly invariant under SL(2,C) transformations
\beq
h_{\alpha} (t) \rightarrow \frac{a \, h_\alpha (t) + b}{c \, h_{\alpha} (t) + d} \label{sl2c} \, ,
\eeq
where $\alpha=1,2$ and $a,b,c,d$ are arbitrary complex numbers.
We will use the Schwinger-Keldysh analysis to derive ordinary differential equations
that are obeyed by $h_{1,2} (t)$ in Section~\ref{sec:largeq}. Naturally, 
these equations will also be invariant under SL(2,C) transformations. 
We will show that for generic initial conditions in the regime $t_{1}<0$ and $t_2 <0$, the solutions
for $h_1 (t_1)$ and $h_2 (t_2)$ at $t_1 > 0$ and $t_2 > 0$ can be written as
\beq
h_1 (t) = \frac{a \, e^{i \theta} e^{\sigma t} + b}{c\, e^{i \theta} e^{\sigma t} + d} 
\quad, \quad h_2 (t) = \frac{a \, e^{-i \theta} e^{\sigma t} + b}{c \, e^{-i \theta} e^{\sigma t} + d} \, . \label{i1}
\eeq
The complex constants $a,b,c,d$, and the real constants $\sigma$, $\theta$ are determined by the
initial conditions in the $t_1 < 0$ and $t_2 < 0$ quadrant of the $t_1$-$t_2$ plane. Note that we pick
a particular SL(2,C) orientation in the $t_1 \leq 0$, $t_2 \leq 0$ quadrant, and then there is no further
SL(2,C) arbitrariness in the $t_1>0$, $t_2 > 0$ quadrant. Inserting Eq.~(\ref{i1}) into Eqs.~(\ref{e2a},\ref{i3})
at $t_1=t_2$, we obtain
\beq
\sigma = 2 \mathcal{J} \sin(\theta) \,.
\eeq

For general $t_1>0$ and $t_2 > 0$, inserting Eq.~(\ref{i1}) into Eq.~(\ref{i3}) we obtain
\beq
g(t_1, t_2 ) = \ln \left[ \frac{- \sigma^2}{4 \mathcal{J}^2 \sinh^2 (\sigma (t_1- t_2)/2 + i \theta)} \right] \,. \label{i4}
\eeq
The surprising feature of this result is that it depends only upon $t=t_1-t_2$, 
and is independent of $\mathcal{T} = (t_1 + t_2)/2$. Indeed Eq.~(\ref{i4}) describes a state in thermal
equilibrium \cite{PG98,Maldacena2016} at an inverse temperature
\beq
\beta_f = \frac{2 (\pi - 2 \theta)}{\sigma} \,. \label{i5}
\eeq
So the large $q$ limit yields a solution in which the fermion Green's function thermalizes instantaneously at $t=0^+$
at the final temperature given by Eq.~(\ref{i5}). 
This could indicate that the 
thermalization rate $\Gamma$ diverges as
at $q \rightarrow \infty$. Alternatively, as pointed out
to us by Aavishkar Patel, the present large $q$ solution
could describe a pre-thermal state, and the $1/q^2$ 
corrections in Eq.~(\ref{e2}) have a finite thermalization rate; in such a scenario, 
thermalization is a two-step process, with the first step occuring much faster than the second. 
But we will not examine the $1/q^2$ corrections here
to settle this issue.

Note also that in the limit $\beta_f \mathcal{J} \gg 1$, we have $\theta \ll 1$ and then
\beq
\sigma = \frac{2 \pi}{\beta_f}\,. \label{e50a}
\eeq
Then Eqs.~(\ref{e2}) and (\ref{i4}) describe the $1/q$ expansion of the low temperature conformal solution \cite{PG98} describing the equilibrium non-Fermi liquid state (see Appendix~\ref{app:conformal}).
And the value of $\sigma$ in Eq.~(\ref{e50a}) is
the maximal Lyapunov exponent for quantum chaos \cite{Maldacena2016a}. This chaos exponent appears in the time evolution 
of $h_{1,2} (t)$. However, in the large $q$ limit, it does not directly control the rapid thermalization rate. We note that recent studies of 
Fermi surfaces coupled to gauge fields, and of disordered metals, found a relaxation/dephasing rate which was larger than the Lyapunov rate \cite{PS17,PCSS17}.

We will begin by setting up the Schwinger-Keldysh formalism for Majorana fermions in the SYK model in Sections~\ref{sec:closed}.
The numerical solution of the Kadanoff-Baym equations
appears in Section~\ref{sec:num}. Finally, the large $q$ limit is described in Section~\ref{sec:largeq}.

\section{Kadanoff-Baym equations from the path integral}
\label{sec:closed}
We construct the path integral for Majorana fermions from the familiar complex fermion path integral~\cite{Sedrakyan2011} by expressing the complex fermions $\Psi_i$ in terms of two real 
fermions $\psi_i$ and $\chi_i$ i.e. $\Psi_i=\psi_i+i\chi_i$. Since 
$\chi_i$ is just a spectator which does not appear in the Hamiltonian, we can disregard its contribution  
and write down the path integral representation of the partition function as
\begin{equation}
Z=\int \mathcal{D}\psi e^{iS[\psi]}
\end{equation}
with the 
action
\begin{equation}
\begin{split}
	S[\psi] &= \int_{\mathcal C} d t \Bigl\{\frac{i}{2} \sum_i \psi_i \partial_t 
\psi_i - i \sum_{i < j} j_{2,ij} f(t) \psi_i \psi_j + \sum_{i<j<k<l} j_{4,ijkl} 
g(t) \psi_i 
\psi_j \psi_k \psi_l\Bigr\} \,.
\label{eq:CTP_Action}
\end{split}
\end{equation}
Here, the Majorana fields $\psi_i$ live on the closed time contour $\mathcal{C}$,
also known as Schwinger-Keldysh contour.
In order to include the disorder, we average the partition function over 
Gaussian distributed couplings as
\begin{equation}
	Z = \int D\psi \int Dj_{2,ij} \int Dj_{4,ijkl} P(j_{4,ijkl}) P(j_{2,ij}) e^{i 
S[\psi]} \, ,
\end{equation}
where the probability distributions are given by
\begin{gather}
\mathcal{P}_{1}(j_{2,ij})=\sqrt{\frac{N}{2J_2^{2}\pi}}\operatorname{exp}
\Big(-\frac { N } { 2J_2^ { 2
}}\sum_{i<j}j_{2,ij}^{2}\Big),\\
\mathcal{P}_{2}(j_{4,ijkl})=\sqrt{\frac{N^{3}}{12J_4^{2}\pi}}\operatorname{exp}
\Big(-\frac {N^{3}}{12J_4^{2}}\sum_{i<j<k<l}j^{2}_{4,ijkl}\Big).
\end{gather}
This realization of disorder allows us to perform the integrals with respect to $j_{2,ij}$ and 
$j_{4,ijkl}$. The Gaussian integrals lead to
\begin{equation}
\begin{split}
 Z = \int D\psi \operatorname{exp}\Bigl\{&-\frac{1}{2}\int_{\mathcal C} dt_1 
\sum_i \psi_i \partial_{t_1} \psi_i - \frac{J_2^2}{4N} \sum_{i,j} \int_{\mathcal C} 
dt_1 \int_{\mathcal C} dt_2 f(t_1) f(t_2)\psi_i(t_1) \psi_i(t_2) \psi_j(t_1) \psi_j(t_2)\\
	&- \frac{3 J_4^2}{4! N^3} \sum_{i,j,k,l} \int_{\mathcal C} dt_1 \int_{\mathcal 
C} dt_2 g(t_1) g(t_2) \psi_i(t_1) \psi_i(t_2) \psi_j(t_1) \psi_j(t_2) \psi_k(t_1) 
\psi_k(t_2) \psi_l(t_1) \psi_l(t_2)\Bigr\}. \nonumber
\end{split} 
\end{equation}
We introduce the bilinear $G$, which must fulfill the constraint
\begin{equation}
	G(t_1, t_2) = - \frac{i}{N}\sum_i \psi_i(t_1) \psi_i(t_2) \,.
\end{equation}
To implement this constraint, we introduce the Lagrange multiplier $\Sigma$ leading to
\begin{equation}
\begin{split}
	Z=&\int D\psi \int DG \int D\Sigma \operatorname{exp}\Bigl\{-\frac{1}{2} 
\int_{\mathcal C} dt_1 \sum_i \psi_i \partial_{t_1} \psi_i + \frac{J_2^2 N}{4} 
\int_{\mathcal C} dt_1 \int_{\mathcal C} dt_2 f(t_1) f(t_2) G(t_1, t_2)^2\\
&-\frac{3 J_4^2 N}{4!} \int_{\mathcal C} dt_1 \int_{\mathcal C} dt_2 g(t_1) 
g(t_2) G(t_1, t_2)^4 + \frac{i}{2} \int_{\mathcal C} dt_1 \int_{\mathcal C} dt_1 
\Sigma(t_1, t_2) \bigl[G(t_1, t_2) + \frac{i}{N} \sum_i \psi_i(t_1) 
\psi_i(t_2)\bigr]\Bigr\}.\nonumber
\end{split}
\end{equation}
Rescaling integration variables as $\Sigma \rightarrow i N\Sigma$ and 
integrating over $\psi$ yields
\begin{equation}
\begin{split}
Z	
	=& \int DG \int D\Sigma \operatorname{exp}\Bigl\{i S\bigl[G, 
\Sigma\bigr]\Bigr\}
\end{split}
\end{equation}
with
\begin{equation}
\begin{split}
	S\bigl[G, \Sigma\bigr] =& -\frac{iN}{2} 
\operatorname{tr}\operatorname{log}\bigl[-i(G_0^{-1} - \Sigma)\bigr] - \frac{i 
J_2^2 N}{4} \int_{\mathcal C} dt_1 \int_{\mathcal C} dt_2 f(t_1) f(t_2) G(t_1, t_2)^2\\
	&+\frac{3i J_4^2 N}{4!} \int_{\mathcal C} dt_1 \int_{\mathcal C} dt_2 g(t_1) 
g(t_2) G(t_1, t_2)^4 + \frac{iN}{2} \int_{\mathcal C} dt_1 \int_{\mathcal C} dt_2 
\Sigma(t_1, t_2) G(t_1, t_2) \label{eq:countouraction}
\end{split}
\end{equation}
and the free Majorana Green's function $G_0^{-1}(t_1,t_2) = i \partial_t \delta_{\mathcal{C}}(t_1,t_2)$.
\!Varying the action with respect to $\Sigma$ and $G$ yields the Dyson equation and the 
 the self-energy, respectively:
\begin{align}
	 G_0^{-1}(t_1, t_2) - G^{-1}(t_1, t_2) &= \Sigma(t_1, t_2) \, , \label{eq:SD} \\
	\Sigma(t_1, t_2) &= J_2^2 f(t_1) f(t_2) G(t_1, t_2) - J_4^2 g(t_1) g(t_2) G(t_1, t_2)^3.
	\label{eq:SelfEnergy_Contour}
\end{align}
Note that the time arguments are with respect to the full time contour, 
and the matrix structure of the Keldysh formalism is implicit. 
We 
define the following two Green's functions  
\begin{subequations}
	\begin{align}
G^{>}(t_1,t_2) &\equiv G(t^{-}_1,t^{+}_2) \, ,	
\label{eq:-+}
	\\
G^{<}(t_1,t_2) &\equiv G(t^{+}_1,t^{-}_2) \, 
\end{align}
\end{subequations}
where $t_i^{-}$ lives on the lower contour and $t_i^{+}$ lives on the upper contour
In the case for Majorana fermions there is only one independent component of the Green's 
function, even in non-equilibrium, due to the relation~\cite{Babadi2015}
\begin{equation}
G^{>}(t_{1}, t_{2})=-G^{<}(t_{2}, t_{1}).
\label{eq:majoranacondition}
\end{equation}
The bare greater and lesser Green's functions are given by
\begin{align}
	G^>_0(t_1, t_2) &= -\frac{i}{2}	\,.
\label{eq:G0_gtr_lss}
\end{align}
We now use $G^{>}$ and $G^{<}$ to obtain the retarded, advanced and Keldysh Green's functions: 
\begin{subequations} 
\begin{align}
\label{eq:retarded}	 G^{R}(t_{1},t_{2})&\equiv\Theta(t_{1}-t_{2})\bigl[G^{>}(t_1, t_2) - G^{<}(t_1, 
t_2)\bigr] \, ,
\\
 \label{eq:advanced}
	 G^{A}(t_{1},t_{2})&\equiv \Theta(t_{2}-t_{1})\bigl[G^{<}(t_1, t_2) - G^{>}(t_1, 
t_2)\bigr] \, ,
\\
 \label{eq:keldysh}
	 G^{K}(t_{1},t_{2})&\equiv G^{>}(t_1, t_2)+G^{<}(t_1, t_2) \,.
	\end{align}
\end{subequations}
Similarly to the Green's function we introduce retarded and advanced self-energies
\begin{align}
\Sigma^R(t_1, t_2) \equiv & \Theta(t_1 - t_2) \bigl[\Sigma^>(t_1, t_2) - \Sigma^<(t_1, t_2)\bigr]\, , \label{eq:SERet}\\
\Sigma^A(t_1, t_2) \equiv & -\Theta(t_2 - t_1) \bigl[\Sigma^>(t_1, t_2) - \Sigma^<(t_1, t_2)\bigr] \label{eq:SEAd}\, 
\end{align}
and for more details on the Schwinger-Keldysh formalism and the saddle point approximation
we refer to  Refs.~\onlinecite{Babadi2015,KamenevBook,Maldacena2016}.
In order to obtain the Kadanoff-Baym equations, we rewrite the Schwinger-Dyson equations \eqref{eq:SD} as
\begin{align}
\int_{\mathcal{C}}dt_3 G_0^{-1}(t_1,t_3)G(t_3,t_2) = \delta_{\mathcal{C}}(t_1,t_2) + \int_{\mathcal{C}}dt_3 \Sigma(t_1,t_3) G(t_3,t_2) \, , \\
\int_{\mathcal{C}}dt_3 G(t_1,t_3)G_0^{-1}(t_3,t_2) = \delta_{\mathcal{C}}(t_1,t_2) + \int_{\mathcal{C}}dt_3  G(t_1,t_3) \Sigma(t_3,t_2) \, .
\end{align}
Using the Langreth rules \cite{stefanucci2013nonequilibrium}
\begin{align}
    \int_{\mathcal{C}}dt_3 \Sigma(t^+_1,t_3) G(t_3,t^+_2) &=  \int_{-\infty}^\infty dt_3 \left\{ \Sigma^R(t_1, t_3) 
G^>(t_3, t_2) + \Sigma^>(t_1, t_2) G^A(t_2, t_3) \right\} \, ,\\
    \int_{\mathcal{C}}dt_3  G(t^+_1,t_3) \Sigma(t_3,t^+_2) &= \int_{-\infty}^\infty dt_3 \left\{  G^R(t_1, t_3) 
\Sigma^>(t_3, t_2) + G^>(t_1, t_2) \Sigma^A(t_2, t_3) \right\} \,,
\end{align}

we obtain the equations of motion for $G^>$ from 
the last equation
\begin{align}
	i \partial_{t_1} G^>(t_1, t_2) &= \int_{-\infty}^\infty dt_3 \Bigl\{\Sigma^R(t_1, t_3) 
G^>(t_3, t_2) + \Sigma^>(t_1, t_3) G^A(t_3, t_2)\Bigr\} \, , \label{eq:kadanoffbaym1} \\
	-i \partial_{t_{2}} G^>(t_1, t_2) &= \int_{-\infty}^\infty dt_3 \Bigl\{G^R(t_1, t_3) 
\Sigma^>(t_3, t_2) + G^>(t_1, t_3) \Sigma^A(t_3, t_2)\Bigr\}, \label{eq:kadanoffbaym2}
\end{align}
where
\begin{gather}
	\Sigma^>(t_1, t_2) = J_2^2 f(t_1) f(t_2) G^>(t_1, t_2) - J_4^2 g(t_1) g(t_2) \bigl(G^>(t_1, 
t_2)\bigr)^3 
\label{eq:selfenergies}
\end{gather}
and $\Sigma^{R}$ and $\Sigma^{A}$ can be obtained from~\eqref{eq:SERet} and ~\eqref{eq:SEAd}. In the following sections we will solve this system of equations numerically and
also solve the generalization of this model in the large q-limit.

\section{Kadanoff-Baym equations: Numerical Study}
\label{sec:num}
In the following we study the non-equilibrium dynamics described by the 
time-dependent Hamiltonian in Eq.~(\ref{eq:Hamiltonian}). The functions $f(t)$ and 
$g(t)$, which are arbitrary so far, specify the quench protocol. We will use them
so switch on(off) or rescale couplings. Hence, all considered quench protocols are of the
form 
\begin{align}
    f(t) &= \alpha_1 \Theta(-t) + \alpha_2 \Theta(t) \, , \nonumber		\\
    g(t) &= \gamma_1 \Theta(-t) + \gamma_2 \Theta(t)	
\end{align}
and will be denoted by $(J_{2,i},J_{4,i}) \rightarrow (J_{2,f},J_{4,f})$. 
Specifically, we choose the following four protocols
\begin{itemize}
	\item[A)] $(J_{2,i},0) \rightarrow (J_{2,f},0)$  
	quench: The energy scale of the random hopping model gets suddenly rescaled.

\item[B)]  $(0,0) \rightarrow (0,J_{4,f})$ 
quench: The system is quenched from an uncorrelated 
state by switching on the interaction of the SYK model.
\item[C)]  $(J_{2,i},0) \rightarrow (0,J_{4,f})$ 
quench: The system is prepared in the ground state of the
random hopping model and then quenched to the SYK Hamiltonian. 
\item[D)]  $(J_{2,i},J_{4,f}) \rightarrow (0,J_4)$ 
quench: Sarting from a ground state with  
the quadratic and the quartic terms present the quadratic term is switched off.
\end{itemize}

The Kadanoff-Baym equations for the SYK model in Eqs.~(\ref{eq:kadanoffbaym1}) and (\ref{eq:kadanoffbaym2})
are be solved numerically. Due to the absence of momentum dependence we are able to explore the long time regime after the quench. The quench time appears at $t_1 = t_2 = 0$. For $t_1 < 0$ and $t_2 < 0$, we are in thermal equilibrium and the time dependence of the Green's function is determined by the Dyson equation. 
When quenching the ground state of the random hopping model, we can use the exact solution for $G^>$ as an initial condition. In all other cases, we solve the 
Dyson equation self-consistently. For further details we refer to Appendix~\ref{App:IC2}.
For $t_1 > 0$ or $t_2 > 0$ we solve Kadanoff-Baym equation. Integrals in the Kadanoff-Baym equation are computed using the trapezoidal rule. After discretizing the integrals the remaining ordinary differential equations are solved using a predictor-corrector scheme, where the corrector is determined self-consistently by iteration. For long times after the quench, the numerical effort is equivalent to a second-order Runge-Kutta scheme because the self-consistency of the predictor-corrector scheme converges very fast. Right after the quench, our approach reduces numerical errors significantly and is thus advantageous.


In order to interpret the long time behavior Green's functions, we briefly comment on
properties of thermal Green's functions. In thermal equilibrium all Green's function only depend on $\tau$.
Further, the imaginary part of the Fourier transformed retarded Green's function is given by the spectral function
\begin{equation}
	A(\omega) = -2 \operatorname{Im} G^R(\omega)\, . 
\end{equation}
In thermal equilibrium the Kubo-Martin-Schwinger (KMS) conditions \cite{KamenevBook} establishes a 
connection between $G^>(\omega)$ and $G^{<}(\omega)$ which is given by
\begin{align}
    G^{>}(\omega) = -e^{\beta \omega} G^{<}(\omega) \,. 
\end{align}
The last relation can be proved by using the Lehmann representation. Employing the definition of $G^R$  
we can extract the spectral function 
\begin{equation}
	A(\omega) =   G^>(\omega) (1 + e^{-\beta \omega})
\end{equation}
from the Green's functions $G^{>}$. 
Similarly the Keldysh component of the fermionic Green's function is related to the spectral 
function,
\begin{equation}
\begin{split}
	iG^K(\omega) &= iG^>(\omega) + iG^<(\omega) = \operatorname{tanh}\bigl(\beta \omega / 2\bigr) A(\omega) \,.
	\label{eq:GK_A}
\end{split}
\end{equation}
Out of equilibrium the Green's functions depend on $\tau$ and $\mathcal{T}$. However, we can still 
consider the Fourier transform with respect to $\tau$ given by
\begin{equation}
	 G^R(\mathcal{T}, \omega) = \int_{0}^{\infty} d\tau \,  e^{-\delta \tau} e^{i\omega \tau}  G^R(\mathcal{T} + \tau/2, \mathcal{T} - \tau/2), \label{eq:GRWigner}
\end{equation}
which is also known as the Wigner transform. From this object we obtain a spectral function out of equilibrium at time $\mathcal{T}$ as
\begin{equation}
	A(\mathcal{T}, \omega) = -2 \operatorname{Im}  G^R(\mathcal{T}, \omega). \label{eq:spectral}
\end{equation}
Numerically, we determine $G^>(t_1, t_2)$  and compute the spectral function abd the Keldysh component. 
In order to investigate thermalization behaviour, we can use a generalization of Eq.~\eqref{eq:GK_A},
\begin{equation}
	\operatorname{tanh}\Bigl(\frac{\beta(\mathcal{T}) \omega}{2}\Bigr) = 
\frac{iG^K(\mathcal{T},\omega)}{A(\mathcal{T}, \omega)},
\label{eq:Thermometer}
\end{equation}
where the quantities on the right hand side are obtained from the Green's functions via a Wigner transformation. In a thermal state this equality holds with a time-independent effective inverse temperature $\beta$. In a non-thermal state, this relation allows one to quantify the deviation from a thermal state and to determine the timescale of thermalization if the final state is indeed thermal.


\subsection{$(J_{2,i},0) \rightarrow (J_{2,f},0)$ quench: Rescaling of the Random-Hopping model}

\begin{figure}
	\centering
	\includegraphics[width=0.7\linewidth]{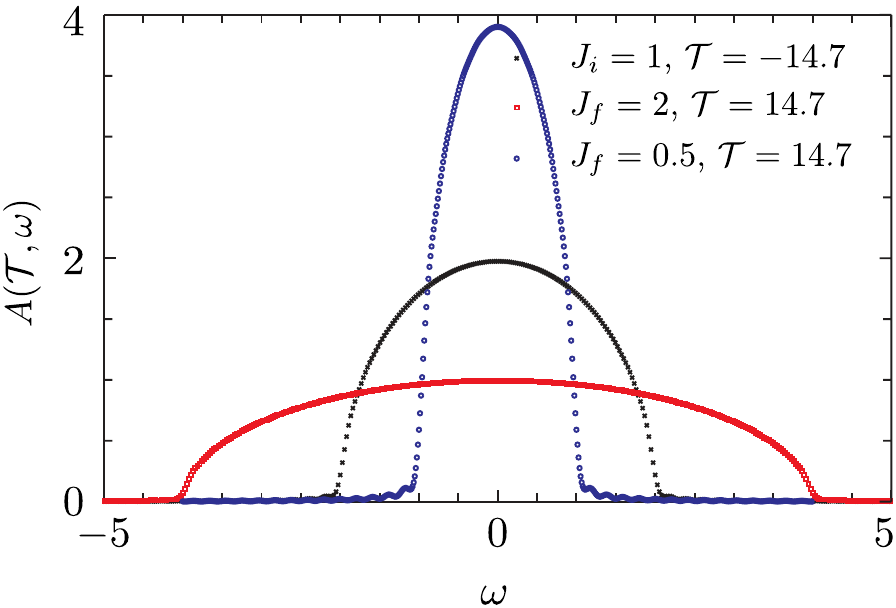}
	\caption{Spectral function of the random hopping model long before ($\mathcal{T} = 
-14.7$) and after ($\mathcal{T} = 14.7$) a parameter quench from $J_{2,i} = 1$ to $J_{2,f} 
= 0.5$ and $2.0$.}
	\label{fig:A_J2_Quench}
\end{figure}
\begin{figure}
	\centering
	\includegraphics[width=0.7\linewidth]{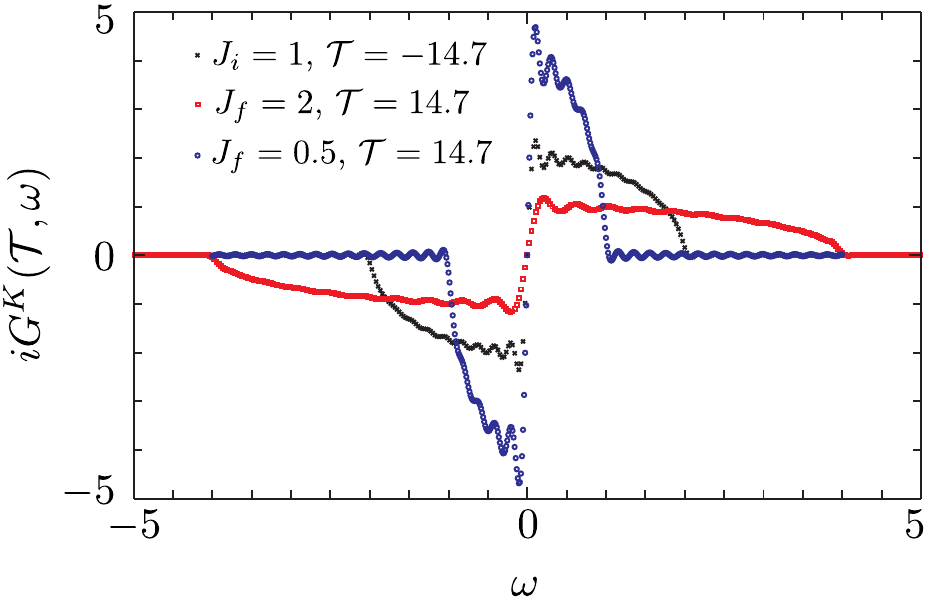}
	\caption{Keldysh component of the Green's function of the random hopping 
model long before ($\mathcal{T} = -14.7$) and after ($\mathcal{T} = 14.7$) a parameter quench from 
$J_{2,i} = 1$ to $J_{2,f} = 0.5$ and $2.0$.}
	\label{fig:iGK_J2_Quench}
\end{figure}
Figure~\ref{fig:A_J2_Quench} shows the spectral function of the random hopping 
model long before and long after a parameter quench. These results were obtained 
from a numerical Fourier transformation of the retarded Green's function as 
described by Eq.~\eqref{eq:GRWigner} with a broadening of $\delta = 0.025$. 
In Fig.~\ref{fig:iGK_J2_Quench}, we show the frequency dependence of the 
Keldysh component of the fermionic Green's function for the same quench 
protocol as in Fig.~\ref{fig:A_J2_Quench}.
\begin{figure}
	\centering
	\includegraphics[width=0.7\linewidth]{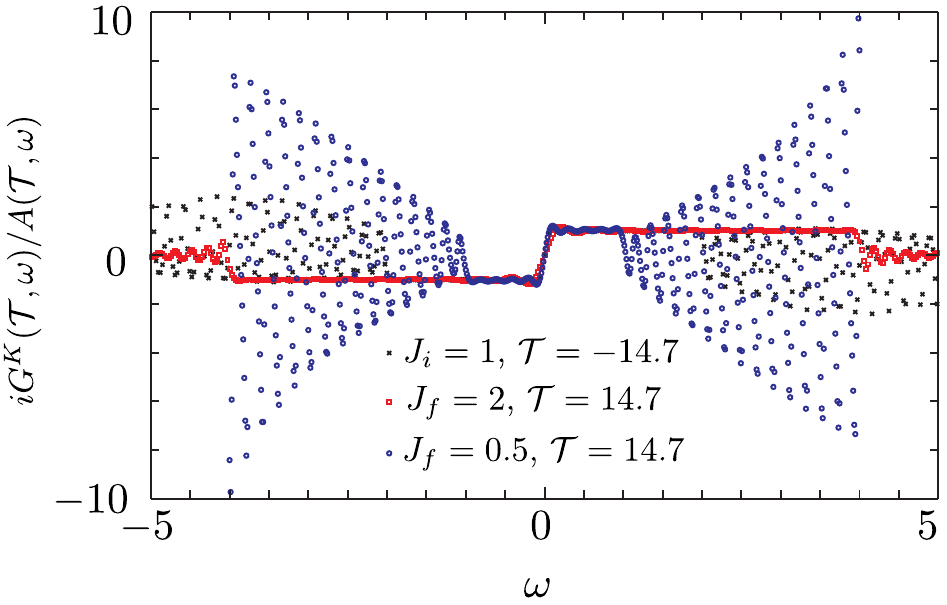}
	\caption{Ratio between the Keldysh component of the Green's function and the 
spectral function of the random hopping model long before ($\mathcal{T} = -14.7$) and 
after ($\mathcal{T} = 14.7$) a parameter quench from $J_{2,i} = 1$ to $J_{2,f} = 0.5$ and 
$2.0$.}
	\label{fig:tanhbw2_J2_Quench}
\end{figure}
In Fig.~\ref{fig:tanhbw2_J2_Quench}, we show the ratio between $i G^K(\mathcal{T}, 
\omega)$ and $A(\mathcal{T}, \omega)$ long before and after the parameter quench. 
The data is only reliable for frequencies $|\omega| < J_{2,f}$. For 
these frequencies, $i G^K(\mathcal{T}, \omega) / A(\mathcal{T}, \omega)$ is almost flat\,--\,up to 
numerical artifacts. We can determine the inverse 
temperature $\beta$ using Eq.~\eqref{eq:Thermometer}, and find that $\beta 
J_2$ is roughly constant during the quench. 
All three results of this $(J_{2,i},0) \rightarrow (J_{2,f},0)$ quench are consistent with a rescaling of energy scales for all quantities. This is expected because the random hopping model has only one energy scale $J_2$,
and the analog of the reparameterization 
in Eq.~(\ref{e1a}) applies here. 
For analytical expressions of the spectral functions we refer to Appendix~\ref{App:IC1}. We further comment,
that whenever an energy scale appears to be one, we measure all other physical quantities with respect to 
this scale.

\subsection{$(0,0) \rightarrow (0,J_4)$ quench: From bare Majoranas to the SYK model}
\begin{figure}
	\centering
	\includegraphics[width=0.7\linewidth]{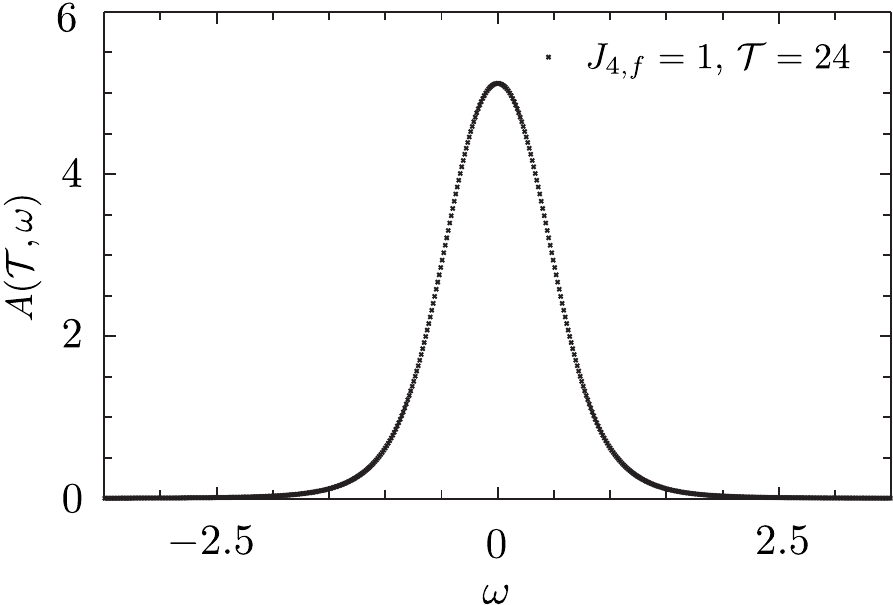}
	\caption{Spectral function of Majorana fermions long after suddenly switching 
on the quartic interaction of the SYK Hamiltonian, with $J_{4,f} = 1$.}
	\label{fig:A_BareMajorana_SYK_Quench}
\end{figure}
In Fig.~\ref{fig:A_BareMajorana_SYK_Quench}, we show the spectral function long 
after suddely switching on the quartic interaction in the SYK model, starting 
from bare, noninteracting Majorana fermions. For an analytical expression of the thermal spectral function we refer to Appendix~\ref{app:conformal}. When interpreting results of this quench protocol, 
it is important to keep in mind that the Hamiltonian before the quench is zero, 
so that any finite $J_{4,f}$ is an arbitrarily strong perturbation.

In this scenario we find that the energy of the system does not change during the quench. It is zero before and after the quench, \ie\ the quench did not pump energy into the system. The Keldysh component of the Green's function vanishes before and after the quench. This is to be expected for free Majorana fermions and is consistent with the temperature of the system being infinite, i.e. $\beta_f = 0$ after the quench.

We interpret this as a result of the particle-hole symmetry of the system 
before the quench when there is only one energy level at $\omega = 0$. The 
quartic interaction broadens this energy level, but does not break the 
particle-hole symmetry. Thus, the spectral weight of the fermions is 
distributed over the entire spectrum. Alternatively, one can argue that the 
Hamiltonian before the quench is zero, so that \emph{any} added interaction is 
in fact infinitely strong.

\subsection{$(J_{2,i},0) \rightarrow (0,J_{4,f})$: From a quadratic to a quartic model}
The quench from the purely quadratic to the purely quartic model decouples the regions $t_1(t_2) < 0$ 
from $t_1(t_2) >0$ as can be seen by inspecting the structure of the Kadanoff-Baym equations and of the self-energy, which reads
\begin{equation}
	\Sigma(t_1, t_2) = \Theta(-t_1) \Theta(-t_2) J_2^2 G(t_1, t_2) - \Theta(t_1) \Theta(t_2) 
J_4^2 G(t_1, t_2)^3
\end{equation}
for this quench protocol. Inserting this into the Kadanoff-Baym equations for 
$t_1>0$ and $t_2>0$, one can see that all time integrals are restricted to 
positive times and the initial condition does not matter: Although $G^>(t_1, t_2)$ 
shows some time evolution for $t_1 \gtrless 0$ and $t_2 \lessgtr 0$ due to 
integrals involving the region with $t_1 < 0$ and $t_2 < 0$, $G^>$ in this 
region does not influence the time evolution at positive $t_1$, $t_2$ because 
$\Sigma(t_1, t_2) = 0$ when $t_1 \gtrless 0$ and $t_2 \lessgtr 0$. Thus, the relevant 
initial condition for the time evolution of $G^>(t_1, t_2)$ is $iG^>(t_1 = 0, t_2 = 
0) = 1/2$. Hence propagating $iG^>$ forward 
in time using the self-energy of the quartic model, we obtain the same time 
evolution at positive times as when starting from bare Majorana fermions.

\subsection{$(J_{2,i},J_{4,i}) \rightarrow (0,J_{4,f})$ 
quench: $J_2$ + $J_4$ model to the SYK model}
\label{sec:j2j4j2}

\begin{figure}
    \centering
    \includegraphics[width=0.75\linewidth]{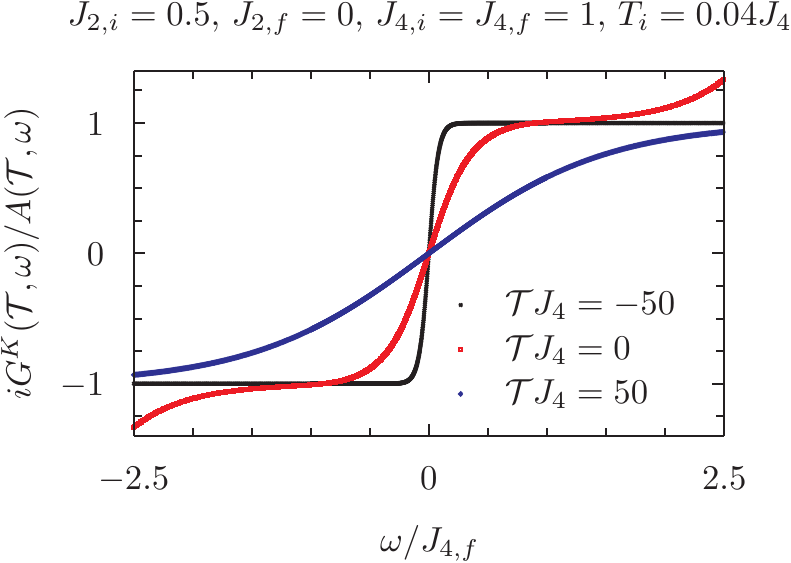}
    \caption{Numerical results of a quench from 
    a $J_2$+$J_4$ model for $t<0$ to a purely $J_4$ model for $t>0$. Fits to this data allow determination of $\beta_{\text{eff}} (\mathcal{T})$.}
    \label{fig:Tanh_J2J4_J4}
\end{figure}
\begin{figure}
    \centering
    \includegraphics[width=0.75\linewidth]{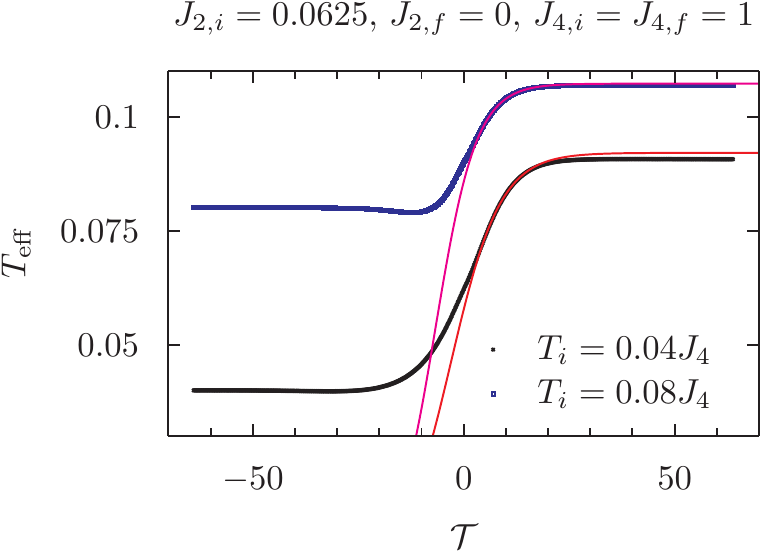}
    \caption{Fits to the values of $\beta_{\text{eff}} (\mathcal{T}) = 1/T_{\text{eff}} (\mathcal{T})$ from results like to those in Fig.~\ref{fig:Tanh_J2J4_J4} to Eq.~(\ref{betaGamma}) allow determination
    of $\Gamma$ for each quench.}
    \label{fig:beta_T}
\end{figure}
\begin{figure}
    \centering
    \includegraphics[width=0.75\linewidth]{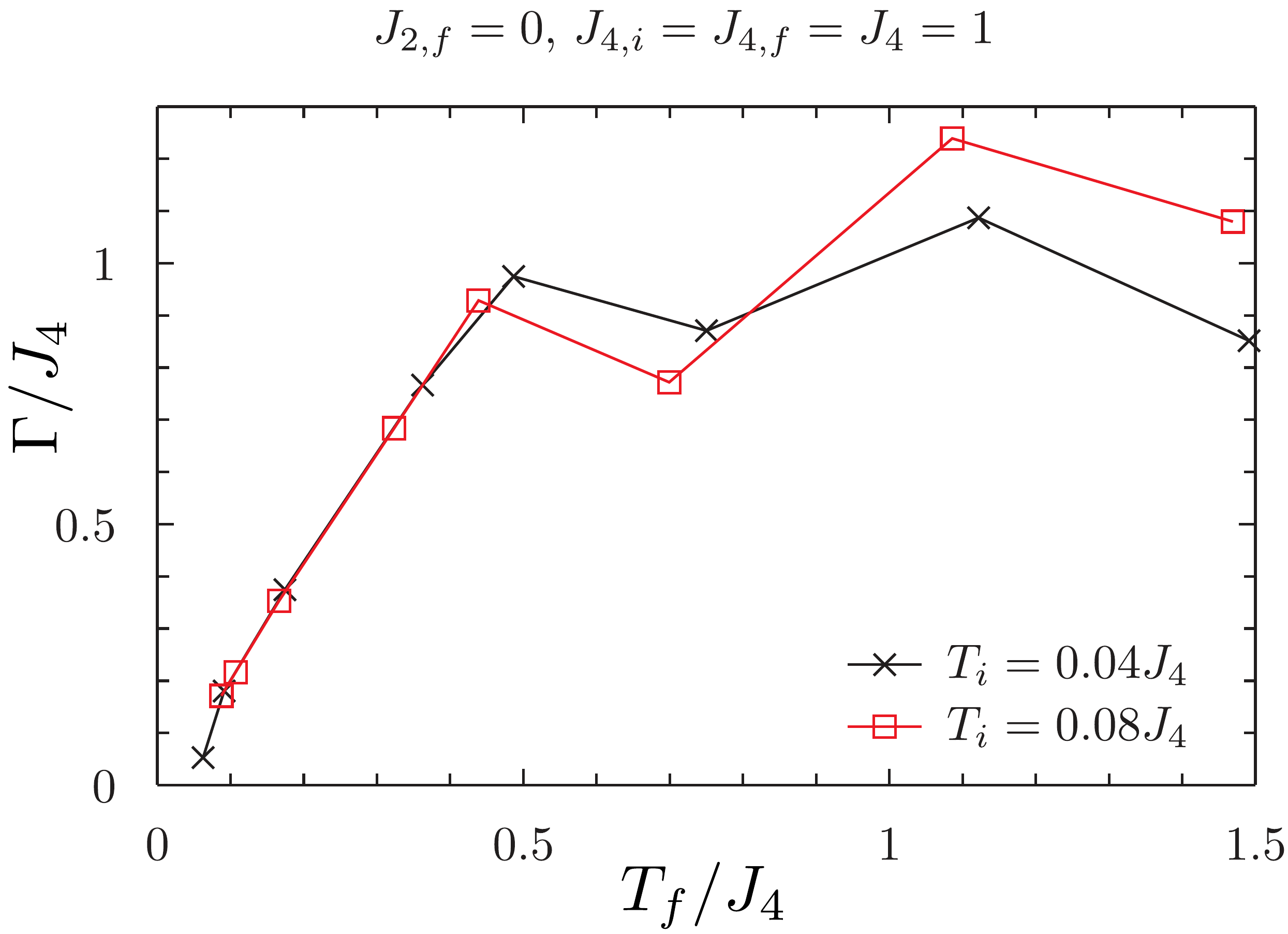}
    \caption{Plots of the values of $\Gamma$ obtained from Fig.~\ref{fig:beta_T} as a function of the final temperature of each quench. Note the proportionality of $\Gamma$ to $T_f$ at small $T_f$.}
    \label{fig:Gamma1}
\end{figure}
In this section, we discuss quenches from a Hamiltonian with a quadratic and a quartic term to a Hamiltonian, where the quadratic term is switched off. Particularly, the time scale of thermalization is the main interest.
We study the validity of the flucutation-dissipation relation and the behavior of the effective temperature $T_\text{eff}(\mathcal T) = \beta_{\text{eff}}^{-1}(\mathcal T)$ in order to detect
thermalization. The effective temperature can either be obtained from the derivative of $i G^K(\mathcal T, \omega) / A(\mathcal T, \omega)$ at $\omega = 0$ or as fit of $\operatorname{tanh}[\omega / (2T_\text{eff})]$ to that function to some frequency interval. 
As shown in Fig.~\ref{fig:Tanh_J2J4_J4}, long after the quench the numerical results for $i G^K(\mathcal T, \omega) / A(\mathcal T, \omega)$ are well described by $\operatorname{tanh}[\omega / (2T_\text{eff})]$, indicating that the system has thermalized. At intermediate times, the low frequency behaviour is still well described by this function. At high frequencies, deviations are visible which could arise due to the fact that $i G^K(\mathcal T, \omega) / A(\mathcal T, \omega)$ is non-thermal. 

In Fig.~\ref{fig:beta_T} we show the time-dependence of the effective temperature and a fit of $T_\text{eff}^{-1}$ to Eq.~(\ref{betaGamma}).
After the quench, the effective temperature shows exponential behavior. In Fig.~\ref{fig:Gamma1} we show the thermalization rate as obtained from such fits. It appears proportional to the final temperature $T_f$ at small final temperatures as noted in Eq.~(\ref{i6}), and saturates at higher temperatures. It is difficult to determine $\Gamma$ numerically at large $T_f/J_4$, and this likely leads to the oscillatory behavior present in Fig.~\ref{fig:Gamma1}.

\section{Kadanoff-Baym equations: Large $q$ limit}
\label{sec:largeq}

This section will
consider a model with a $q$ fermion coupling $J (t)$, and a $pq$ fermion coupling $J_p(t)$. For now, we keep the time-dependence of both
couplings arbitrary. We will find that the initial state is exactly solvable for $p=2$ and $p=1/2$.

The Kadanoff-Baym equations in Eqs.~(\ref{eq:kadanoffbaym1}), (\ref{eq:kadanoffbaym2}) and (\ref{eq:selfenergies})
for $G^> (t_1, t_2)$ now become
\begin{align}
i \frac{\partial}{\partial t_1} G^>(t_1,t_2) &= - i^q \int_{-\infty}^{t_1} dt_3 \, J(t_1) J(t_3) \left[ (G^>)^{q-1} (t_1,t_3) -(G^<)^{q-1} (t_1,t_3) \right] G^{>}(t_3, t_2)  \notag \\
&+  i^q \int_{-\infty}^{t_2} dt_3 \, J(t_1) J(t_3) (G^>)^{q-1} (t_1,t_3) 
\left[ G^>(t_3,t_2) -G^<(t_3,t_2) \right]   \notag \\
& - i^{pq} \int_{-\infty}^{t_1} dt_3 \, J_p (t_1) J_p (t_3) \left[ (G^>)^{pq-1} (t_1,t_3) -(G^<)^{pq-1} (t_1,t_3) \right] G^{>}(t_3, t_2)  \notag \\
&+  i^{pq} \int_{-\infty}^{t_2} dt_3 \, J_p (t_1) J_p (t_3) (G^>)^{pq-1} (t_1,t_3) 
\left[ G^>(t_3,t_2) -G^<(t_3,t_2) \right] \, ,  \notag \\
-i \frac{\partial}{ \partial t_2} G^>(t_1,t_2) &= - i^q \int_{-\infty}^{t_1} dt_3 \, J(t_3) J(t_2) \left[ G^>(t_1,t_3) - G^<(t_1,t_3) \right] (G^{>})^{q-1} (t_3,t_2)  \notag \\
&+ i^q \int_{-\infty}^{t_2} dt_3 \, J(t_3) J(t_2) G^>(t_1,t_3) \left[ (G^>)^{q-1} (t_3,t_2) -(G^<)^{q-1} (t_3,t_2) \right]
\notag \\
& - i^{pq} \int_{-\infty}^{t_1} dt_3 \, J_p(t_3) J_p(t_2) \left[ G^>(t_1,t_3) - G^<(t_1,t_3) \right] (G^{>})^{pq-1} (t_3,t_2)  \label{e1} \\
&+ i^{pq} \int_{-\infty}^{t_2} dt_3 \, J_p(t_3) J_p(t_2) G^>(t_1,t_3) \left[ (G^>)^{pq-1} (t_3,t_2) -(G^<)^{pq-1} (t_3,t_2) \right]
\,. \notag
\end{align}
where we have defined
\beq
J_p (t) \equiv J_p f(t) \quad , \quad J(t) \equiv J f(t)\,.
\eeq 
Also, recall that Eq.~(\ref{eq:majoranacondition}) relates $G^>$ to
$G^<$.

We note a property of the Kadanoff-Baym equations in Eq.~(\ref{e1}), connected to a comment in Section~\ref{sec:intro} above Eq.~(\ref{eq:wigner}). 
If we set $J_p (t) = 0$, then all dependence of Eq.~(\ref{e1})
on $J(t)$ can be scaled away by reparameterizing time via
\beq
\int^t J(t') dt' \rightarrow t \,. \label{e1a}
\eeq
This implies that correlations remain in thermal equilibrium in the new time co-ordinate.
However, when both $J_p (t)$ and $J (t)$ are non-zero, such a reparameterization is
not sufficient, and there is non-trivial quench
dynamics, as was shown by our numerical study in Section~\ref{sec:num}. Below, we will see that
the quench dynamics can also be trivial in the limit $q \rightarrow \infty$, even when both
$J_p(t)$ and $J(t)$ are both non-zero. But this result arises from fairly non-trivial computations which are described below, and in particular from an SL(2,C) symmetry of the 
parameterization of the equations. 

In the large $q$ limit, we assume a solution of the form in Eq.~(\ref{e2}).
Inserting (\ref{e2}) into (\ref{e1}), we obtain to leading order in $1/q$
\bea
\frac{\partial}{\partial t_1} g(t_1,t_2) &=& 
2 \int_{-\infty}^{t_2} dt_3  \, \mathcal{J}(t_1) \mathcal{J}(t_3) e^{g(t_1,t_3)} 
 -\int_{-\infty}^{t_1} dt_3 \, \mathcal{J}(t_1) \mathcal{J}(t_3) \left[ e^{g(t_1,t_3)} + e^{g(t_3,t_1)} \right]  
  \nn
&+& 2 \int_{-\infty}^{t_2} dt_3  \, \mathcal{J}_p (t_1) \mathcal{J}_p (t_3) e^{pg(t_1,t_3)} -\int_{-\infty}^{t_1} dt_3 \, \mathcal{J}_p (t_1) \mathcal{J}_p (t_3) \left[ e^{pg(t_1,t_3)} + e^{pg(t_3,t_1)} \right]  \, ,
  \nn
 \frac{\partial}{ \partial t_2} g(t_1,t_2) &=& 2 \int_{-\infty}^{t_1} dt_3  \, \mathcal{J}(t_3) \mathcal{J}(t_2) e^{g(t_3,t_2)} 
-  \int_{-\infty}^{t_2} dt_3  \, \mathcal{J}(t_3) \mathcal{J}(t_2) \left[ e^{g(t_3,t_2)} + e^{g (t_2,t_3)} \right] \label{e3} \\
&+& 2 \int_{-\infty}^{t_1} dt_3  \, \mathcal{J}_p(t_3) \mathcal{J}_p(t_2) e^{pg(t_3,t_2)} 
-  \int_{-\infty}^{t_2} dt_3  \, \mathcal{J}_p(t_3) \mathcal{J}_p (t_2) \left[ e^{pg(t_3,t_2)} + e^{pg (t_2,t_3)} \right] \, ,
\nonumber
\eea
where
\beq
\mathcal{J}^2 (t) = q J^2 (t) 2^{1-q} \quad , \quad \mathcal{J}_p^2 (t) = q J_p^2 (t) 2^{1-pq}\label{e4}
\eeq
It is a remarkable fact that these non-linear, partial, 
integro-differential equations are exactly solvable for our quench protocol and all initial conditions, as we will show in the remainder of this section. The final exact solution appears in Section~\ref{sec:exact}, and surprisingly shows that the solution is instantaneously in thermal equilibrium at $t=0^+$.

Taking the derivatives of either equation in Eq.~(\ref{e3}) we obtain
\beq
\frac{\partial^2}{\partial t_1 \partial t_2} g(t_1, t_2) = 2 \mathcal{J}(t_1) \mathcal{J} (t_2) e^{g (t_1, t_2)} + 2 \mathcal{J}_p(t_1) 
\mathcal{J}_p (t_2) e^{pg (t_1, t_2)} \,.
\label{e5}
\eeq
We will look at the case where $\mathcal{J}_p (t)$ is non-zero only for $t<0$, while $\mathcal{J}(t)$ is time independent:
\beq
\mathcal{J}_p (t) = \mathcal{J}_p \, \Theta(-t) \quad, \quad  \mathcal{J} (t) = \mathcal{J} \,.\label{impulse}
\eeq
\begin{figure}
\begin{center}
\includegraphics[width=5in]{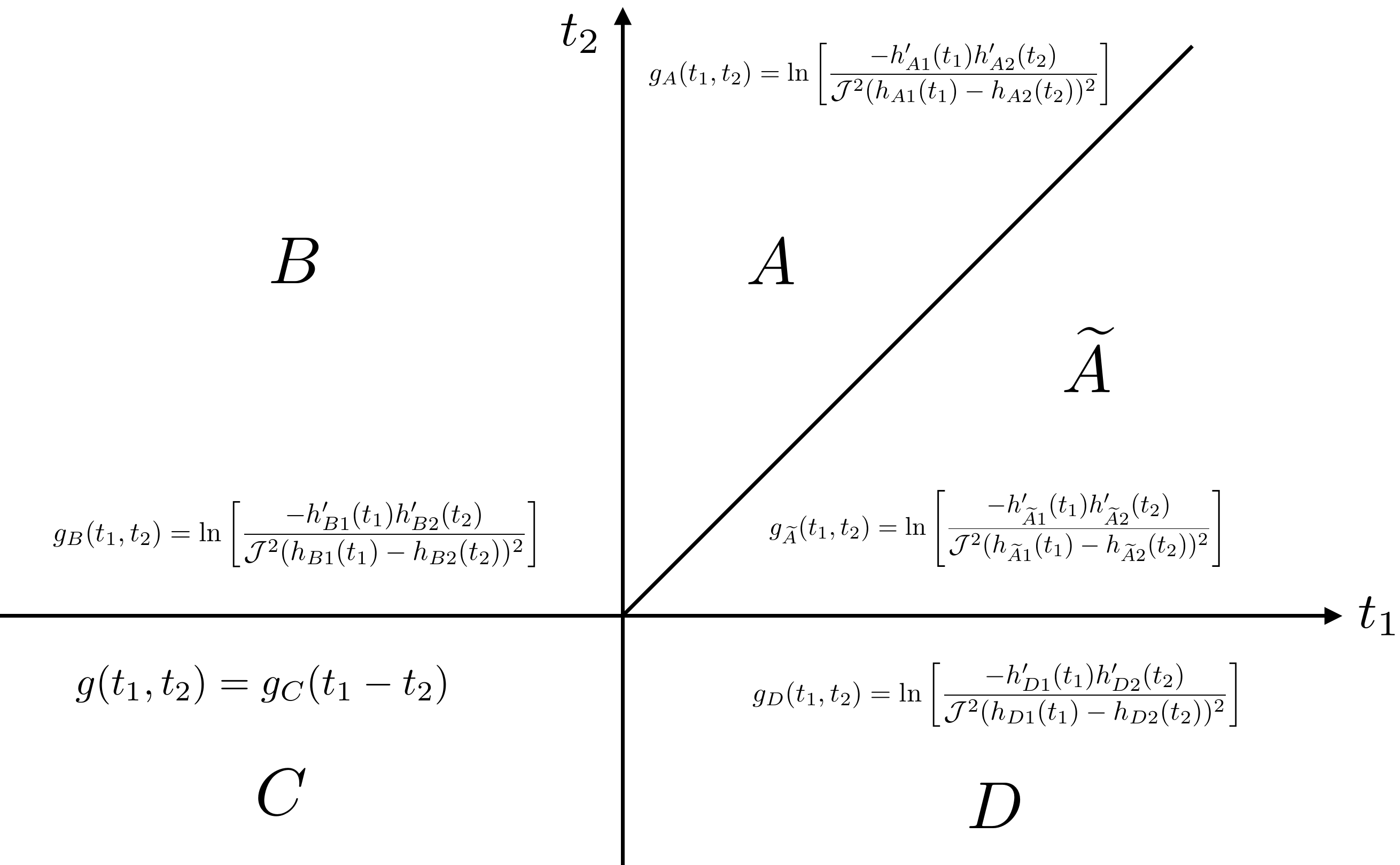}
\end{center}
\caption{The regions in the $t_1$-$t_2$ plane.}
\label{fig:quadrants}
\end{figure}
Then Eq.~(\ref{e5}) is the Lorentzian Liouville equation in three of the four quadrants of the $t_1$-$t_2$ plane (labeled as in Fig.~\ref{fig:quadrants}), and its most general solution in these quadrants is 
\beq
g_\alpha (t_1, t_2) = 
\ln \left[ \frac{-h_{\alpha 1}' (t_1) h_{\alpha 2}' (t_2)}{\mathcal{J}^2 (h_{\alpha 1} (t_1) - h_{\alpha 2} (t_2))^2} \right] \quad, \quad \mbox{for $\alpha = A,\widetilde{A},B,D$} \,.
\label{eh}
\eeq
Eq.~(\ref{eh}) will also apply for $\alpha = C$ when $\mathcal{J}_p=0$.
Eq.~(\ref{e2a}) implies that
the functions $h_{A 1} (t)$ and $h_{A 2} (t)$ obey
\beq
h_{A 1}' (t) h_{A 2}' (t) = - \mathcal{J}^2 (h_{A 1} (t) - h_{A 2} (t))^2 \,, \label{eh2}
\eeq
and similarly
\beq
h_{\widetilde{A} 1}' (t) h_{\widetilde{A} 2}' (t) = - \mathcal{J}^2  (h_{\widetilde{A} 1} (t) - h_{\widetilde{A} 2} (t))^2 \,, \label{eh2t}
\eeq
Note that $g_\alpha (t_1, t_2)$ remains invariant under the SL(2,C) mapping in Eq.~(\ref{sl2c}).
Given the symmetries of Eq.~(\ref{e3}) and the thermal initial conditions, we look for solutions which obey
\beq
g(t_2, t_1) = \left[g(t_1, t_2)\right]^\ast \,. \label{ggs}
\eeq
This property can be related to the causality of the Kadanoff-Baym equations as is explained in 
Appendix~\ref{App:ConProperty}.
We can satisfy Eq.~(\ref{ggs}) in quadrants A, B, D by the relations
\bea
h_{\widetilde{A}1} (t) &=& h_{A2}^\ast (t) \, , \nn
h_{\widetilde{A}2} (t) &=& h_{A1}^\ast (t) \, ,  \nn 
h_{D1} (t) &=& h_{B2}^\ast (t) \, ,  \nn
h_{D2} (t) &=& h_{B1}^\ast (t) \, .
\label{eh20}
\eea

The system is in equilibrium in quadrant C, and so the Green's function depends only upon time differences.
We describe the nature of this equilibrium solution in Appendix~\ref{app:quadrantC}.
In the following subsections, we describe the results of inserting the parameterization in Eq.~(\ref{eh})
back into Eq.~(\ref{e3}) to obtain ordinary differential equations for $h_{\alpha 1}$ and $h_{\alpha 2}$.

\subsection{Quadrant B}

From Eq.~(\ref{e3}), we obtain for $t_1 < 0$ and $t_2 > 0$
\bea
-\frac{\partial}{\partial t_1} g_B (t_1,t_2) &=&   -\mathcal{J}^2 \int_{-\infty}^{\infty} dt' \,   e^{g_C (t')} \,  \mbox{sgn} (t'-t_1)
 - 2  \mathcal{J}^2 \int_{0}^{t_2} dt_3  \,  e^{g_B (t_1,t_3)}  \nn
 &~&~~~~~ - \mathcal{J}_p^2 \int_{-\infty}^{\infty} dt' \mbox{sgn}(t'-t_1) e^{p g_C (t')} \, ,\nn
  \frac{\partial}{ \partial t_2} g_B (t_1,t_2) &=& 2\mathcal{J}^2   \int_{-\infty}^{t_1} dt_3  \, e^{g_B (t_3,t_2)} 
-  \mathcal{J}^2  \int_{-\infty}^{0} dt_3  \,  \left[ e^{g_B (t_3,t_2)} + e^{g_D (t_2,t_3)} \right] \nn &~&~~~~~
-  \mathcal{J}^2  \int_{0}^{t_2} dt_3  \,  \left[ e^{g_A (t_3,t_2)} + e^{g_{\widetilde{A}} (t_2,t_3)} \right]\,. \label{eomB}
\eea
Inserting Eq.~(\ref{eh}) in Eq.~(\ref{eomB}) we obtain
\bea
- \frac{h_{B1}'' (t_1)}{h_{B1}' (t_1)} &=&  -\mathcal{J}^2 \int_{-\infty}^{\infty} dt' \,   e^{g_C (t')} \,  \mbox{sgn} (t'-t_1) - \frac{2 h_{B1}' (t_1)}{h_{B1} (t_1) - h_{B2} (0)}  \nn
 &~&~~~~~ - \mathcal{J}_p^2 \int_{-\infty}^{\infty} dt' \mbox{sgn}(t'-t_1) e^{p g_C (t')} \,, \label{eomB2} \\
\frac{h_{B2}'' (t_2)}{h_{B2}' (t_2)} &=&  - \frac{2 h_{B2}' (t_2)}{h_{B1} (-\infty) - h_{B2} (t_2)} \label{eomB3a}
\\ &~&
- \frac{h_{B2}' (t_2)}{h_{B1} (0) - h_{B2} (t_2)}
+ \frac{h_{B2}' (t_2)}{h_{B1} (-\infty) - h_{B2} (t_2)}
- \frac{h_{B2}^{\ast\prime} (t_2)}{h_{B1}^\ast (0) - h_{B2}^\ast (t_2)}
+ \frac{h_{B2}^{\ast\prime} (t_2)}{h_{B1}^\ast (-\infty) - h_{B2}^\ast (t_2)} \nn
&~& - \frac{h_{A2}^{\prime} (t_2)}{h_{A1} (t_2) - h_{A2} (t_2)}
+ \frac{h_{A2}^{\prime} (t_2)}{h_{A1} (0) - h_{A2} (t_2)}
- \frac{h_{A2}^{\ast\prime} (t_2)}{h_{A1}^\ast (t_2) - h_{A2}^\ast (t_2)}
+ \frac{h_{A2}^{\ast\prime} (t_2)}{h_{A1}^\ast (0) - h_{A2}^\ast (t_2)} \,.
\nonumber
\eea

We also have the compatibility condition at the boundary between regions B and C, which is
\beq
g_C (t_1) = \ln \left[ \frac{-h_{B 1}' (t_1) h_{B 2}' (0)}{\mathcal{J}^2 (h_{B 1} (t_1) - h_{B 2} (0))^2} \right] \,; \label{eomB3}
\eeq
taking the derivative of this equation we obtain precisely the first equation in Eq.~(\ref{eomB2}) after using Eq.~(\ref{eomC}). 
This reassures us that the system of equations are not overdetermined. We can easily integrate Eq.~(\ref{eomB3}) to obtain
\beq
\frac{h_{B2}' (0)}{h_{B1} (t_1) - h_{B2} (0)} - \frac{h_{B2}' (0)}{h_{B1} (-\infty) - h_{B2} (0)} = 
\mathcal{J}^2 \int_{-\infty}^{t_1} dt \, e^{g_C (t)} \,. \label{exact4}
\eeq

\subsection{Region A}

Now $t_2 >  t_1 > 0$. 
From the second Eq.~(\ref{e3}) we obtain
\bea
  \frac{\partial}{ \partial t_2} g_A (t_1,t_2) &=& 2\mathcal{J}^2   \int_{-\infty}^{0} dt_3  \, e^{g_B (t_3,t_2)} 
  + 2\mathcal{J}^2   \int_{0}^{t_1} dt_3  \, e^{g_A (t_3,t_2)}
-  \mathcal{J}^2  \int_{-\infty}^{0} dt_3  \,  \left[ e^{g_B (t_3,t_2)} + e^{g_D (t_2,t_3)} \right] \nn &~&~~~~~
-  \mathcal{J}^2  \int_{0}^{t_2} dt_3  \,  \left[ e^{g_A (t_3,t_2)} + e^{g_{\widetilde{A}} (t_2,t_3)} \right]\,. \label{eomA}
\eea
Inserting Eq.~(\ref{eh}) into Eq.~(\ref{eomA}), we obtain
\bea
\frac{h_{A2}^{\prime\prime} (t_2)}{h_{A2}^{\prime} (t_2)} &=&  \frac{2 h_{B2}' (t_2)}{h_{B1} (0) - h_{B2} (t_2)} - \frac{2 h_{B2}' (t_2)}{h_{B1} (-\infty) - h_{B2} (t_2)} - \frac{2 h_{A2}^{\prime} (t_2)}{h_{A1} (0) - h_{A2} (t_2)} \label{eomA5}
\\ &~&
- \frac{h_{B2}' (t_2)}{h_{B1} (0) - h_{B2} (t_2)}
+ \frac{h_{B2}' (t_2)}{h_{B1} (-\infty) - h_{B2} (t_2)}
- \frac{h_{B2}^{\ast\prime} (t_2)}{h_{B1}^\ast (0) - h_{B2}^\ast (t_2)}
+ \frac{h_{B2}^{\ast\prime} (t_2)}{h_{B1}^\ast (-\infty) - h_{B2}^\ast (t_2)} \nn
&~& - \frac{h_{A2}^{\prime} (t_2)}{h_{A1} (t_2) - h_{A2} (t_2)}
+ \frac{h_{A2}^{\prime} (t_2)}{h_{A1} (0) - h_{A2} (t_2)}
- \frac{h_{A2}^{\ast\prime} (t_2)}{h_{A1}^\ast (t_2) - h_{A2}^\ast (t_2)}
+ \frac{h_{A2}^{\ast\prime} (t_2)}{h_{A1}^\ast (0) - h_{A2}^\ast (t_2)} \,.
\nonumber
\eea

From the first Eq.~(\ref{e3}) we obtain
\bea
-\frac{\partial}{\partial t_1} g_A (t_1,t_2) &=&   -\mathcal{J}^2 \int_{-\infty}^{0} dt_3 \,   e^{g_D (t_1, t_3)} 
 -\mathcal{J}^2 \int_{0}^{t_1} dt_3 \,   e^{g_{\widetilde{A}} (t_1, t_3)} 
  - 2\mathcal{J}^2 \int_{t_1}^{t_2} dt_3 \,   e^{g_A (t_1, t_3)} \nn
  &~&~  +\mathcal{J}^2 \int_{-\infty}^{0} dt_3 \,   e^{g_B (t_3, t_1)} 
   + \mathcal{J}^2 \int_{0}^{t_1} dt_3 \,   e^{g_A (t_3, t_1)} \,.\label{eomA2}
\eea
Inserting Eq.~(\ref{eh}) into Eq.~(\ref{eomA2}), we obtain
\bea
-\frac{h_{A1}^{\prime\prime} (t_1)}{h_{A1}^{\prime} (t_1)} &=& 
 \frac{2 h_{A1}^{\prime} (t_1)}{h_{A2} (t_1) - h_{A1} (t_1)} \label{eomA6} \\
&~&~
 - \frac{h_{B2}^{\ast\prime} (t_1)}{h_{B1}^{\ast} (0) - h_{B2}^{\ast} (t_1)}
+ \frac{h_{B2}^{\ast\prime} (t_1)}{h_{B1}^{\ast} (-\infty) - h_{B2}^{\ast} (t_1)}
+ \frac{h_{B2}^{\prime} (t_1)}{h_{B1} (0) - h_{B2} (t_1)} -  \frac{h_{B2}^{\prime} (t_1)}{h_{B1} (-\infty) - h_{B2} (t_1)} \nn
&~&~
+ \frac{h_{A2}^{\prime} (t_1)}{h_{A1} (t_1) - h_{A2} (t_1)} - \frac{h_{A2}^{\prime} (t_1)}{h_{A1} (0) - h_{A2} (t_1)}
- \frac{h_{A2}^{\ast\prime} (t_1)}{h_{A1}^{\ast} (t_1) - h_{A2}^\ast (t_1)} + \frac{h_{A2}^{\ast\prime} (t_1)}{h_{A1}^\ast (0) - h_{A2}^\ast (t_1)} \,.
\nonumber
\eea
From Eqs.~(\ref{eomA5}) and (\ref{eomA6}) we obtain
\beq
\frac{h_{A1}^{\prime\prime} (t)}{h_{A1}^{\prime} (t)} + \frac{h_{A2}^{\prime\prime} (t)}{h_{A2}^{\prime} (t)} = 
2 \left(\frac{h_{A1}'(t)- h_{A2}' (t) }{h_{A1} (t)  - h_{A2} (t)} \right)\,.
\eeq
This is precisely the logarithmic derivative of the Majorana condition in Eq.~(\ref{eh2}).
The compatibility condition at the boundary of region A and region B is
\beq
\frac{h_{A1}' (0) h_{A2}^{\prime} (t_2)}{(h_{A1} (0) - h_{A2} (t_2))^2} = \frac{h_{B1}' (0) h_{B2}' (t_2)}{(h_{B1} (0) - h_{B2} (t_2))^2} \,. \label{sA1}
\eeq
This can be integrated to 
\beq
\frac{h_{A1}' (0)}{h_{A2} (t_2) - h_{A1} (0)} = \frac{h_{B1}' (0)}{h_{B2} (t_2)-h_{B1} (0) } + c_4\,, \label{sA2}
\eeq
where $c_4$ is a constant of integration.

\subsection{Combined equations}

We adopt a simple choice to solve Eq.~(\ref{sA2})
\bea
h_{B2} (t) &=& h_{A2} (t)\,, \nn
h_{B1} (0) &=& h_{A1} (0)\,, \nn
h_{B1}' (0) &=& h_{A1}' (0)\,. \label{exact3}
\eea
Then we find that Eqs.~(\ref{eomB3a}) and (\ref{eomA5}) are consistent with each other.
Collecting all equations, we need to solve
\begin{align}
-\frac{h_{A1}^{\prime\prime} (t)}{h_{A1}^{\prime} (t)} &= 
- \frac{2 h_{A1}^{\prime} (t)}{h_{A1} (t) - h_{A2} (t)} 
+ \frac{h_{A2}^{\ast\prime} (t)}{h_{B1}^{\ast} (-\infty) - h_{A2}^{\ast} (t)}
-  \frac{h_{A2}^{\prime} (t)}{h_{B1} (-\infty) - h_{A2} (t)} \nn &
+ \frac{h_{A2}^{\prime} (t)}{h_{A1} (t) - h_{A2} (t)} 
- \frac{h_{A2}^{\ast\prime} (t)}{h_{A1}^{\ast} (t) - h_{A2}^\ast (t)} \quad , \quad t \geq 0 \label{efinal1} \\
h_{A 1}' (t) h_{A 2}' (t) &= - \mathcal{J}^2  (h_{A 1} (t) - h_{A 2} (t))^2 \quad , \quad t \geq 0 \label{efinal2} \\
\frac{h_{B2}' (0)}{h_{B1} (t) - h_{B2} (0)} &= \frac{h_{B2}' (0)}{h_{B1} (-\infty) - h_{B2} (0)} + 
\mathcal{J}^2 \int_{-\infty}^{t} dt' \, e^{g_C (t')} \quad , \quad t \leq 0 \label{efinal3} \\
h_{A1} (0) & = h_{B1} (0) \,,\label{efinal4} \\
h_{A1}' (0) & = h_{B1}' (0) \,,\label{efinal5} \\
h_{A2} (0) & = h_{B2} (0) \,.\label{efinal6}
\end{align}
It can be verified that all expressions above are invariant SL(2,C) transformations of the $h_{\alpha 1}$
and $h_{\alpha 2}$ fields.
Given the values of $h_{B1}(-\infty)$, $h_{B2} (0)$, $h_{B2}' (0)$, and $g_C (t)$, 
Eqs.~(\ref{efinal3}-\ref{efinal6}) determine the values of $h_{A1} (0)$,  $h_{A1}' (0)$, and $h_{A2} (0)$.
Then
Eqs.~(\ref{efinal1},\ref{efinal2}) uniquely determine $h_{A1} (t)$ and $h_{A2} (t)$ for all $t \geq 0$. 
Because of the SL(2,C) invariance, the values chosen for $h_{B1}(-\infty)$, $h_{B2} (0)$, $h_{B2}' (0)$ won't matter for the 
final result for $g_A (t_1, t_2)$. 

\subsection{Exact solution}
\label{sec:exact}

A solution of the form in Eq.~(\ref{Cexact}) does not apply to Eqs.~(\ref{efinal1}-\ref{efinal6}) in region A because it does
not have enough free parameters to satisfy the initial conditions. However, we can use the SL(2,C) invariance of Eqs.~(\ref{efinal1}-\ref{efinal3})
to propose the ansatz similar to that in Eq.~(\ref{i1}) (here, we redefine constants by
factors of $e^{i \theta}$):
\begin{align}
h_{A1} (t) = \frac{ a \, e^{\sigma t} + c}{c \, e^{\sigma t} + d} \quad, \quad h_{A2} (t) = \frac{ a \, e^{-2 i \theta}  e^{ \sigma t} + b}{ c \, e^{-2 i \theta} e^{\sigma t} + d} \,. \label{exact1}
\end{align}
We can now verify that Eq.~(\ref{exact1}) is an exact solution of Eqs.~(\ref{efinal1}-\ref{efinal6}). This solution is characterized
by 4 complex numbers $a$, $b$, $c$, $d$ and two real numbers $\theta$, $\sigma$. These are uniquely determined 
from the values of $h_{B1} (-\infty)$, $h_{A1} (0)$, $h_{A1}' (0)$, and $h_{A2} (0)$ 
by the solution of the following 6 equations
\begin{align}
a d - b c & = 1 \,, \nn
\sigma &= 2 \mathcal{J} \sin(\theta) \,,\nn
e^{-4 i \theta} &= \frac{(b - d \, h_{B1} (- \infty))(a^\ast - c^\ast \, h_{B1}^\ast (-\infty))}{(b^\ast - d^\ast \, h_{B1}^\ast (- \infty))(a - c \, h_{B1} (-\infty))} \,, \nn
h_{A1} (0) &= \frac{ a + b}{c + d} \,, \nn
h_{A2} (0) &= \frac{ a \, e^{-2 i \theta} + b }{c \, e^{-2 i \theta} + d } \,, \nn
h_{A1}' (0) &= \frac{2 \sin(\theta)}{(c + d)^2} \,.
\label{exact2}
\end{align}
It is now easy to verify that $g_A$ takes the form in Eq.~(\ref{i4}), and so all of quadrant A is also in thermal 
equilibrium. We also numerically integrated Eqs.~(\ref{efinal1}-\ref{efinal6}), starting from generic 
initial conditions, and verified that the numerical solution obeyed the 
expressions in Eqs.~(\ref{exact1}) and (\ref{exact2}).

The above solution determines the values of $\sigma$ and $\theta$ in the final state, and hence the value of the final temperature via Eq.~(\ref{i5}). In general, this will be different from the value of the initial temperature in quadrant C.

Eqs.~(\ref{exact1}) and (\ref{exact2}) also determine the solutions in the other quadrants via expressions specified earlier. The solutions in 
region $\widetilde{A}$ follow from the conjugacy property in Eq.~(\ref{eh20}).
In quadrant B, we have $h_{B2} (t) = h_{A2} (t)$ in Eq.~(\ref{exact3}), while 
$h_{B1} (t)$ was specified in Eq.~(\ref{exact4}) using the initial state in 
quadrant C. Note that the solution in quadrant B is not of a thermal form, as $h_{B1}$ and $h_{B2}$ are
not simply related as in Eq.~(\ref{exact1}).
The solution in quadrant D follows from that in quadrant B via
the conjugacy property in Eq.~(\ref{eh20}). And, finally, the initial
state in quadrant C was described in Appendix~\ref{app:quadrantC}.

\section{Conclusions}
\label{sec:conc}

Quantum many-body systems without quasiparticle excitations are expected to locally thermalize
in the fastest possible times of order $\hbar/(k_B T)$ as $T \rightarrow 0$, where $T$ is the absolute temperature of the final
state \cite{ssbook}. 
This excludes {\it e.g.\/} the existence of systems in which the local thermalization rate, $\Gamma \sim T^{p}$ as $T \rightarrow 0$
with $p<1$, and no counterexamples have been found.

In this paper we examined SYK models, which are systems which saturate the more rigorous bounds
on the Lyapunov time to reach quantum chaos \cite{Maldacena2016a}. Our numerical study of the model with a final Hamiltonian with $q=4$
showed that this system does thermalize rapidly, and the thermalization rate is consistent with 
$\Gamma = C T$ at low $T$ where
$C$ is dimensionless constant,
as indicated in Eq.~(\ref{i6}) and Fig.~\ref{fig:Gamma1}.

We also studied a large $q$ limit of the SYK models, where an exact analytic solution of the non-equilibrium
dynamics was possible. Here we found that thermalization of the fermion Green's function was instantaneous. It will be necessary to study higher order corrections in $1/q$ to understand
how this connects to the numerical $q=4$ numerical 
solution: does the constant $C \rightarrow \infty$
as $q \rightarrow \infty$, or (as pointed to us
by Aavishkar Patel) is thermalization at large $q$ a two-step process. In two-step scenario, a very rapid pre-thermalization (which we have computed) is followed by a slower true thermalization of higher order corrections.

Finally, we comment on a remarkable feature of the large $q$ solution given by Eq.~(\ref{i3}) and (\ref{i1}):
its connection with the Schwarzian.
The Schwarzian was proposed as an effective Lagrangian for the low energy limit of the equilibrium theory.
Specifically, consider the Euler-Lagrange equation of motion of a Lagrangian, $\mathcal{L}$, which is the Schwarzian
of $h(t)$
\beq
\mathcal{L}[h(t)] = \frac{h^{\prime\prime\prime} (t)}{h^{\prime} (t)} - \frac{3}{2} \left( \frac{h^{\prime\prime} (t)}{h^{\prime} (t)}\right)^2 \,.
\eeq
The equation of motion is
\beq
\left[ h^{\prime} (t) \right]^2 h^{\prime\prime\prime\prime} (t) 
+ 3 \left[ h^{\prime\prime} (t) \right]^3 
- 4 h^{\prime} (t) h^{\prime\prime} (t)  h^{\prime\prime\prime} (t) = 0 \,. \label{i10}
\eeq
It can now be verified that the expressions for $h_{1,2} (t)$ in Eq.~(\ref{i1}) (and Eq.~(\ref{exact1})) both obey Eq.~(\ref{i10}). We note, however, that we did not obtain Eq.~(\ref{i1}) by the solution of 
Eq.~(\ref{i10}): instead, Eq.~(\ref{i1}) was obtained by the solution of the Schwinger-Keldysh
equations of the large-$q$ SYK model in Eq.~(\ref{e3}).
This connection with the Schwarzian indicates that gravitational models \cite{SS10,SS10b,AAJP15,kitaev2015talk,Maldacena2016,KJ16,HV16} of the quantum quench in AdS$_2$, which map to a Schwarzian boundary theory,
exhibit instant thermalization as in the large $q$ limit. 

Indeed, the equation of motion of the metric in two-dimensional gravity \cite{AAJP15} takes a form
identical to that for the two-point fermion correlator in Eq.~(\ref{liouville}). And
studies of black hole formation
in AdS$_2$ from a collapsing shell of matter 
show that the Hawking temperature of the black hole jumps instantaneously
to a new equilibrium value after the passage of the shell \cite{HV16,engelsoythesis}. 
These features are strikingly similar to 
those obtained in our large $q$ analysis. There have been studies of quantum quenches in AdS$_2$, either in the context of
quantum impurity problems \cite{Erdmenger17}, or in the context of higher-dimensional black holes which
have an AdS$_2$ factor in the low energy limit \cite{Withers16}; it would be useful to analytically extract the behavior of just AdS$_2$ by extending such studies.

We thank J. Maldacena of informing us about another work \cite{IKJM17} which studied aspects
of thermalization of SYK models by very different methods, along with connections to gravity on AdS$_2$.

\section*{Acknowledgments} 
We would like to thank Wenbo Fu, Daniel Jafferis, Michael Knapp, Juan Maldacena, Ipsita Mandal, Thomas Mertens, Robert Myers, Aavishkar Patel, Stephen Shenker, and Herman Verlinde 
for valuable discussions. 
JS was supported by the National Science Foundation Graduate Research 
Fellowship under Grant No. DGE1144152.
AE acknowledges support from the German National Academy of Sciences 
Leopoldina through grant LPDS~2014-13. AE would like to thank 
the Erwin Schrödinger International Institute for Mathematics and Physics in 
Vienna, Austria, for hospitality and financial support during the workshop on 
"Synergies between Mathematical and Computational Approaches to Quantum 
Many-Body Physics". VK acknowledges support from the Alexander von Humboldt Foundation through a Feodor Lynen Fellowship. This research was supported by the NSF under Grant 
DMR-1360789 and MURI grant W911NF-14-1-0003 from ARO. Research at Perimeter 
Institute is supported by the Government of Canada through Industry Canada and 
by the Province of Ontario through the Ministry of Economic Development \& 
Innovation. SS also acknowledges support from Cenovus Energy at Perimeter 
Institute.

\appendix

\section{Spectral function and Green's functions of random hopping model \label{App:IC1}} 

The random hopping model of Majorana fermions, \ie the SYK Hamiltonian for $q 
= 2$, can be solved exactly~\cite{Maldacena2016}. We repeat the calculation 
within our notation in the following. The Matsubara Green's function follows 
from the solution of
\begin{equation}
	G(i\omega_n)^{-1} = i\omega_n - \Sigma(i\omega_n) = i\omega_n + J_2^2 
G(i\omega_n).
\end{equation}
This can be rewritten as a quadratic equation for 
$\bigl(G(i\omega_n)\bigr)^{-1}$ and matching the branches with the correct 
high-frequency asymptotics and symmetry yields the propagator
\begin{equation}
	G(i\omega_n) = \frac{2}{i\omega_n + i \operatorname{sgn}(\omega_n) 
\sqrt{\omega_n^2 + 4 J_2^2}}.
\end{equation}
Analytically continuing to the real frequency axis from the upper half plane 
yields the retarded Green's function,
\begin{equation}
	G^R(\omega+i\delta) = \frac{2}{\omega+i\delta+i\sqrt{4J_2^2 - (\omega 
+i\delta)^2}},
\end{equation}
where $\delta$ can be set to zero due to the presence of the finite imaginary 
part. Note that we exploited that $\operatorname{sgn}\omega_n \rightarrow 1$ in 
the analytic continuation from the upper half plane. We obtain
\begin{equation}
	A(\omega) = -2 \operatorname{Im} G^R(\omega+i\delta) = 
\frac{2}{J_2}\sqrt{1 - \Bigl(\frac{\omega}{2J_2}\Bigr)^2}\qquad \text{for } 
|\omega| < 2 J_2,
\label{eq:A_q2}
\end{equation}
which is the well-known semicircular density of states due to random hopping.

 This yields
\begin{equation}
	i G^>(t) = \frac{1}{2J_2 t} \bigl(J_1(2 J_2 t) - i H_1(2 J_2 t)\bigr),
	\label{eq:Ggtr_q2}
\end{equation}
where $J_1$ and $H_1$ are the Bessel function of the first kind (\verb!BesselJ! 
in \verb!Mathematica!) and the Struve function (\verb!StruveH! in 
\verb!Mathematica!), respectively.

\section{Spectral Function in the Conformal Limit}
\label{app:conformal}

In the scaling limit at non-zero temperature the  retarded Green's function is 
given by the following expression
\begin{equation}
iG^{R}(t)=2b(\cos\pi\Delta)\left(\frac{\pi}{\beta\sinh\frac{\pi t}{\beta}}\right)^{2\Delta}\theta(t)\,,
\end{equation}
where $\Delta = 1/q$ is the fermion scaling dimension.
At $q=4$ we obtain 
\begin{equation}
iG^{R}_{c}(t)=\sqrt{2}b\left(\frac{\pi}{\beta\sinh\frac{\pi t}{\beta}}\right)^{\frac{1}{2}}\theta(t) \label{eq:retardedconformal}
\end{equation}
The Wigner transform of the retarded Green's functions is given by 
\begin{align}
	iG^{R}(\omega) &= \sqrt{2} b  \int_0^{\infty}dt e^{i\omega t} \left(\frac{\pi}{\beta\sinh\frac{\pi t}{\beta}}\right)^{\frac{1}{2}} \notag \\
	&= 
	b  \left(\frac{\pi}{\beta}\right)^{-\frac{1}{2}}  B\left( \frac{1}{2}; \frac{1}{4} -\frac{i \beta\omega}{2 \pi}\right)\,,
	\label{eq:retardedconformalfourier}
\end{align}
where $b=({4\pi J^{2}_{4}})^{-{1}/{4}}$. The associated spectral function is
\begin{equation}
A(\mathcal{T},\omega)=2b \left(\frac{\pi}{\beta}\right)^{-\frac{1}{2}} \text{Re}\left[B\left( \frac{1}{2}; \frac{1}{4} -\frac{i \beta\omega}{2 \pi}\right)\right] \,.
\end{equation}

\section{Details on the Numerical Solution of Kadanoff-Baym equation \label{App:IC2}}
The Green's function is typically determined on two-dimensional grids in $(t,t')$ space with $8000 \times 8000$ or $12000 \times 12000$ points, where the quench happens after half of the points in each direction. When starting from initial states in which only $J_2$ is finite, the Green's function decays algebraically in time. This leads to significant finite size effects in Fourier transforms. In order to reduce the latter, most numerical results were obtained by starting from a thermal state in which $J_2$ and $J_4$, or only $J_4$, are finite. In these cases, the Green's function decays exponentially as a function of the relative time dependence.

We checked the quality and consistency of the results by monitoring  the conservation of
energy, the normalization of the spectral function and the real and imaginary part of the retarded propagator are 
Kramers-Kronig consistent for long times after the quench. 

In order to time-evolve the Kadanoff-Baym equations we have to determine $G^{<}(t_1,t_2)$ for $t_1,t_2$. When quenching the ground state of the random hopping model, we can use the exact solution for $G^>$ as initial condition. When we do not have an analytical expression the Green's function we solve the 
Dyson equation self-consistently according to the following scheme:
\begin{enumerate}
	\item Prepare $i G^>$ with an initial guess, for example the propagator of 
the random hopping model.
	\item \label{Dyson:Iter} Computation of retarded self-energy in time domain:
	\begin{equation}
		i\Sigma^R(t) = \Theta(t)(i \Sigma^>(t) + i\Sigma^>(-t))
	\end{equation}
	\item Fourier transformation
	\begin{equation}
		i\Sigma^R(\omega) = \int_{-\infty}^\infty dt e^{i \omega t} i\Sigma^R(t)
	\end{equation}
	\item Dyson equation:
	\begin{equation}
		G^R(\omega) = \frac{1}{\omega - \Sigma^R(\omega)}
	\end{equation}
	\item Determine spectral function
	\begin{equation}
		A(\omega) = -2 \operatorname{Im} G^R(\omega)
	\end{equation}
	\item Determine $i G^>(\omega)$ from spectral function,
	\begin{equation}
		i G^>(\omega) = (1 - n_F(\omega)) A(\omega)
	\end{equation}
	Note that this is the only step in the self-consistency procedure where 
the temperature $\beta^{-1}$ enters through the Fermi function $n_F$.
	\item Fourier transformation to time domain
	\begin{equation}
		iG^>(t) = \int_{-\infty}^\infty \frac{d\omega}{2\pi} e^{-i \omega t} 
iG^>(\omega)
	\end{equation}
	\item Continue with step~\ref{Dyson:Iter} until convergence is reached.
\end{enumerate}

\section{Conjugacy property of $g(t_1,t_2)$ \label{App:ConProperty}}
\begin{figure}[htb!]
\begin{center}
\includegraphics[width=3in]{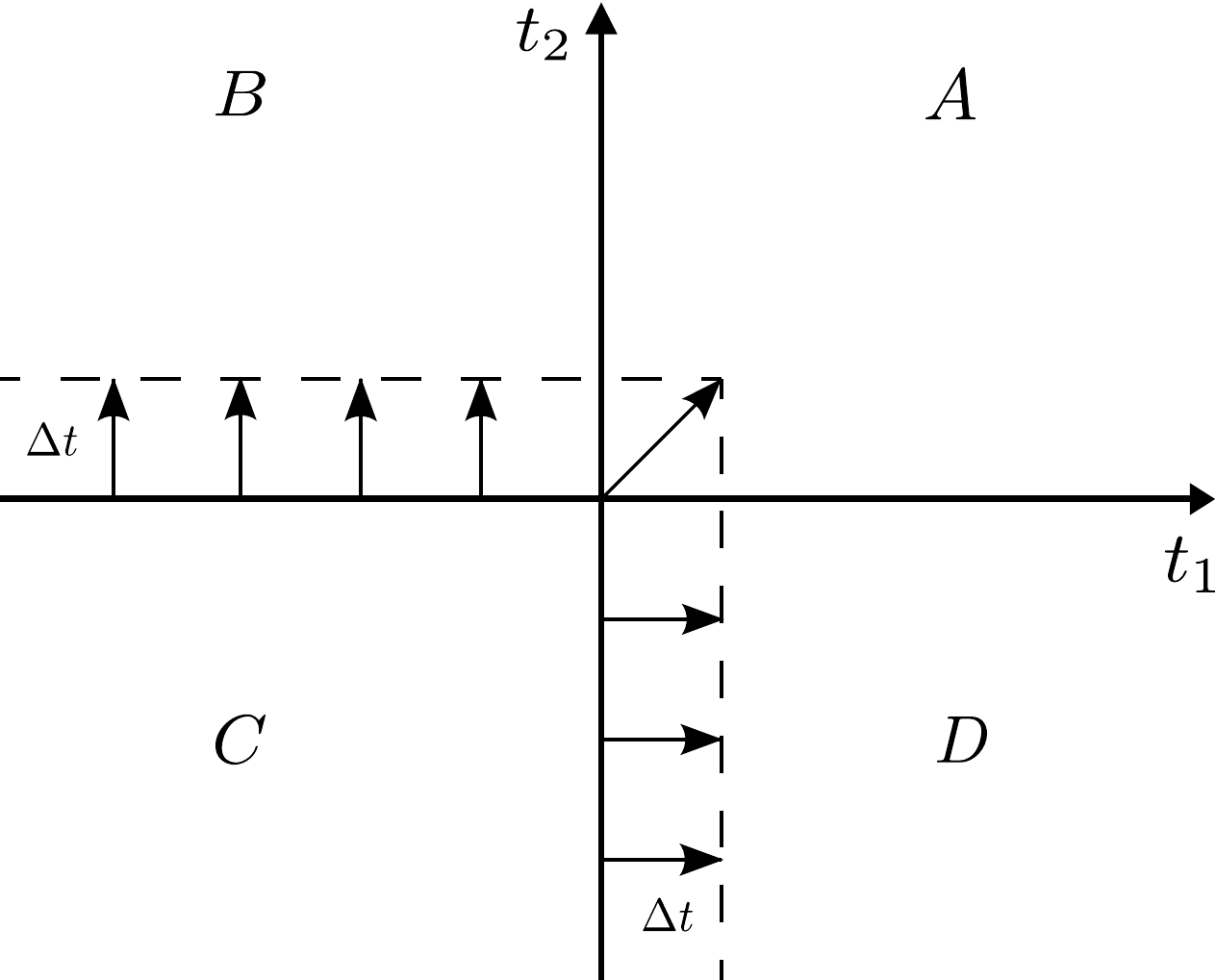}
\end{center}
\caption{The propagation of the conjugacy property in the $t_1$-$t_2$ plane.}
\label{fig:quadrantsApp}
\end{figure}

In this appendix we illustrate that the causality structure of the Kadanoff-Baym equations eq. \eqref{e3}  leads to  the property 
\begin{align}
[g(t_1,t_2)]^{\ast} = g(t_2,t_1) 
\end{align}
in all four quadrants of Fig.~\ref{fig:quadrantsApp}. Since the system is thermal in quadrant $C$
the conjugacy property $[g(t_1,t_2)]^{\ast} = g(t_2,t_1)$ can be read off the thermal solution for $t_1,t_2\leq0$.
Next, we consider the propagation from the line $\left\{(t_1,t_2)\in \mathbb{R}^2 \,| t_2 = 0, t_1\leq 0 \right\}$ for an infinitesimal time $\Delta t$ in the $t_2$  
direction to the line $\left\{(t_1,t_2)\in \mathbb{R}^2 \,| t_2 = \Delta t, t_1\leq 0 \right\}$. 
We discretize equation \eqref{e3} as follows
\begin{align}
\frac{1}{\Delta t} \left[ g(t_1,\Delta t) - g(t_1,0) \right] &= 2 \int_{-\infty}^{t_1} dt_3  \, \mathcal{J}(t_3) \mathcal{J}(0) e^{g(t_3,0)} 
-  \int_{-\infty}^{0} dt_3  \, \mathcal{J}(t_3) \mathcal{J}(0) \left[ e^{g(t_3,0)} + e^{g (0,t_3)} \right] \nn
&+ 2 \int_{-\infty}^{t_1} dt_3  \, \mathcal{J}_p(t_3) \mathcal{J}_p(0) e^{pg(t_3,0)} 
-  \int_{-\infty}^{0} dt_3  \, \mathcal{J}_p(t_3) \mathcal{J}_p (0) \left[ e^{pg(t_3,0)} + e^{pg (0,t_3)} \right]  
\end{align}	
and  solve the resulting equation for $g(t_1,\Delta t)$. Taking the complex conjugate leads to
\begin{align}
 g(t_1,\Delta t)^{\ast} &=g(t_1,0)^{\ast} + \Delta t \,  \Bigg\{  2 \int_{-\infty}^{t_1} dt_3  \, \mathcal{J}(t_3) \mathcal{J}(0) e^{g(t_3,0)^{\ast}} 
-  \int_{-\infty}^{0} dt_3  \, \mathcal{J}(t_3) \mathcal{J}(0) \left[ e^{g(t_3,0)^{\ast}} + e^{g (0,t_3)^{\ast}} \right] \nn
&+ 2 \int_{-\infty}^{t_1} dt_3  \, \mathcal{J}_p(t_3) \mathcal{J}_p(0) e^{pg(t_3,0)^{\ast}} 
-  \int_{-\infty}^{0} dt_3  \, \mathcal{J}_p(t_3) \mathcal{J}_p (0) \left[ e^{pg(t_3,0)^{\ast}} + e^{pg (0,t_3)^{\ast}} \right]  \Bigg\}\,.
\label{e6}
\end{align}
On the right hand side we are allowed to use the property $g(t_1,t_2)^{\ast} = g(t_2,t_1)$
since all $g$'s are still living in the $C$ quadrant.
We obtain
\begin{align}
 g(t_1,\Delta t)^{\ast} &=g(0,t_1) + \Delta t \,  \Bigg\{  2 \int_{-\infty}^{t_1} dt_3  \, \mathcal{J}(t_3) \mathcal{J}(0) e^{g(0,t_3)} 
-  \int_{-\infty}^{0} dt_3  \, \mathcal{J}(t_3) \mathcal{J}(0) \left[ e^{g(0,t_3)} + e^{g (t_3,0)} \right] \nn
&+ 2 \int_{-\infty}^{t_1} dt_3  \, \mathcal{J}_p(t_3) \mathcal{J}_p(0) e^{pg(0,t_3)} 
-  \int_{-\infty}^{0} dt_3  \, \mathcal{J}_p(t_3) \mathcal{J}_p (0) \left[ e^{pg(0,t_3)} + e^{pg (t_3,0)} \right]  \Bigg\}\,.
\end{align}
Next we let the system propagate from the line $\left\{(t_1,t_2)\in \mathbb{R}^2 \,| t_2 \leq 0, t_1 = 0 \right\}$ for an infinitesimal time $\Delta t$ in the $t_1$  
direction to the line $\left\{(t_1,t_2)\in \mathbb{R}^2 \,| t_1 = \Delta t, t_2\leq 0 \right\}$. 
We discretize equation \eqref{e1} as follows
\begin{align}
-\frac{1}{\Delta t} \left[ g(\Delta t,t_2) - g(0,t_2) \right] &=  \int_{-\infty}^{0} dt_3 \, \mathcal{J}(0) \mathcal{J}(t_3) \left[ e^{g(0,t_3)} + e^{g(t_3,0)} \right]  
- 2 \int_{-\infty}^{t_2} dt_3  \, \mathcal{J}(0) \mathcal{J}(t_3) e^{g(0,t_3)}   \nn
&+  \int_{-\infty}^{0} dt_3 \, \mathcal{J}_p (0) \mathcal{J}_p (t_3) \left[ e^{pg(0,t_3)} + e^{pg(t_3,0)} \right]  
- 2 \int_{-\infty}^{t_2} dt_3  \, \mathcal{J}_p (0) \mathcal{J}_p (t_3) e^{pg(0,t_3)}\,
\end{align}	
We set $t_2=t_1$ in the last equation and solve for $g(\Delta t,t_1)$ leading to
\begin{align}
g(\Delta t,t_1)  &=  g(0,t_1) -\Delta t \, \Bigg\{  \int_{-\infty}^{0} dt_3 \, \mathcal{J}(0) \mathcal{J}(t_3) \left[ e^{g(0,t_3)} + e^{g(t_3,0)} \right]  
- 2 \int_{-\infty}^{t_1} dt_3  \, \mathcal{J}(0) \mathcal{J}(t_3) e^{g(0,t_3)}   \nn
&+  \int_{-\infty}^{0} dt_3 \, \mathcal{J}_p (0) \mathcal{J}_p (t_3) \left[ e^{pg(0,t_3)} + e^{pg(t_3,0)} \right]  
- 2 \int_{-\infty}^{t_1} dt_3  \, \mathcal{J}_p (0) \mathcal{J}_p (t_3) e^{pg(0,t_3)} \Bigg\} \,. \label{e9}
\end{align}	
Comparing the right hand side of \eqref{e6} with \eqref{e9} we conclude $g(t_1,\Delta t)^{\ast} = g(\Delta t, t_1)$. Furthermore, the point $g(\Delta t, \Delta t)$ fulfills the 
conjugate property trivially. In total we propagated the conjugate property one time slice.
Repeating this argument for every time slice of size $\Delta t$, the property will hold in all four 
quadrants of Fig. \ref{fig:quadrantsApp}.

\section{Initial state in the large $q$ limit}
\label{app:quadrantC}

In quadrant C in Fig.~\ref{fig:quadrants}, all Green's functions are dependent only on time differences, and and then we can write 
\beq
g_C (t_1, t_2) \equiv g_C (t_1 - t_2)
\eeq
and Eq.~(\ref{e3}) as 
\beq
\frac{d g_C (t)}{d t} =  \mathcal{J}^2 \int_{-\infty}^{\infty} dt' \,   e^{g_C (t')} \,  \mbox{sgn} (t'-t)
+\mathcal{J}_p^2 \int_{-\infty}^{\infty} dt' \,   e^{pg_C (t')} \,  \mbox{sgn} (t'-t) \, . \label{eomC}
\eeq
This implies the second order differential equation (also obtainable from Eq.~(\ref{e5}))
\beq
-\frac{d^2 g_C}{dt^2} = 2 \mathcal{J}^2 e^{g_C} + 2 \mathcal{J}_p^2 
e^{pg_C} \, .
\label{e50}
\eeq

Eq.~(\ref{e50}) turns out to be exactly solvable at $p=2$ (pointed out to us by Wenbo Fu, following \cite{cenke2017}).
We write $g_C (t) = \ln (-1/f(t))$ and then Eq.~(\ref{e50}) becomes
\beq
\frac{1}{f} \frac{d^2 f}{dt^2} - \frac{1}{f^2} \left( \frac{df}{dt} \right)^2 = - \frac{2 \mathcal{J}^2}{f} + \frac{2 \mathcal{J}^2_2}{f^2}
\eeq
The solution of this differential equation yields
\beq
g_C (t) = \ln \left[ \frac{-\sigma^2}{\sqrt{4 \mathcal{J}^4 + 2 \mathcal{J}_2^2 \sigma^2} \cosh ( \sigma t - 2 i \theta) - 2 \mathcal{J}^2} \right] \,, \label{sp2}
\eeq
with
\beq
\cos(2 \theta) = \frac{2 \mathcal{J}^2 - \sigma^2}{\sqrt{ 4 \mathcal{J}^4 + 2 \mathcal{J}_2^2 \sigma^2}} \,.
\eeq
By evaluating the Fourier transform of the Green's function we find $G^> (-\omega) = e^{-\beta_i \omega} G^> (\omega)$, or by analytically continuing to imaginary time, we find that the
initial inverse temperature is
\beq
\beta_i = \frac{2 (\pi - 2 \theta)}{\sigma} \,. \label{temp}
\eeq
We will ultimately be interested in the scaling limit in which $\theta \rightarrow 0$ and $\sigma \ll \mathcal{J}, \mathcal{J}_2$.

Eq.~(\ref{e50}) is also exactly soluble at $p=1/2$ by
\beq
g_C (t) = 2 \ln \left[ \frac{\sigma^2/2}{i \sqrt{4 \mathcal{J}_{1/2}^4 +  \mathcal{J}^2 \sigma^2} \sinh ( \sigma t/2 -  i \theta) + 2\mathcal{J}_{1/2}^2} \right] \,, \label{sp3}
\eeq
where now
\beq
\sin(\theta) = \frac{\sigma^2/2 - 2\mathcal{J}_{1/2}^2  }{\sqrt{  4\mathcal{J}_{1/2}^4 + \mathcal{J}^2 \sigma^2}} \,.
\eeq
The value of the inverse initial temperature remains as in Eq.~(\ref{temp}).

Note that both solutions in Eqs.~(\ref{sp2}) and (\ref{sp3}) obey
\beq
g_C (-t) = g_C^\ast (t)
\eeq

It is also useful to recast the solution in quadrant C for the case $\mathcal{J}_p =0$ in the form of Eq.~(\ref{eh}).
We subdivide quadrant C into two subregions just as in quadrant A.
From Eqs.~(\ref{e3}) and (\ref{eh}) we obtain for $t_2 > t_1$ 
\bea
\frac{h_{C1}^{\prime\prime} (t_1)}{h_{C1}^{\prime} (t_1)} &=& 
\frac{2 h_{C2}^{\ast\prime} (t_1)}{h_{C1}^\ast (t_1) - h_{C2}^\ast (t_1)} 
-\frac{2 h_{C2}^{\ast\prime} (t_1)}{h_{C1}^\ast (-\infty) - h_{C2}^\ast (t_1)} - \frac{2 h_{C1}^{\prime} (t_1)}{h_{C2} (t_1) - h_{C1} (t_1)} \nn
 &~& - \frac{h_{C2}^{\ast\prime} (t_1)}{h_{C1}^\ast (t_1) - h_{C2}^\ast (t_1)} + \frac{h_{C2}^{\ast\prime} (t_1)}{h_{C1}^\ast (-\infty) - h_{C2}^\ast (t_1)} - \frac{h_{C2}^{\prime} (t_1)}{h_{C1} (t_1) - h_{C2} (t_1)} + \frac{h_{C2}^{\prime} (t_1)}{h_{C1} (-\infty) - h_{C2} (t_1)} \nn
\frac{h_{C2}^{\prime\prime} (t_2)}{h_{C2}^{\prime} (t_2)} &=& -\frac{2 h_{C2}^{\prime} (t_2)}{h_{C1} (-\infty) - h_{C2} (t_2)} 
-\frac{h_{C2}^{\prime} (t_2)}{h_{C1} (t_2) - h_{C2} (t_2)} + \frac{h_{C2}^{\prime} (t_2)}{h_{C1} (-\infty) - h_{C2} (t_2)} \nn
&~& - \frac{h_{C2}^{\ast\prime} (t_2)}{h_{C1}^\ast (t_2) - h_{C2}^\ast (t_2)} + \frac{h_{C2}^{\ast\prime} (t_2)}{h_{C1}^\ast (-\infty) - h_{C2}^\ast (t_2)} \,. \label{eC1}
\eea
Adding the equations in Eq.~(\ref{eC1}), we have
\beq
\frac{h_{C1}^{\prime\prime} (t)}{h_{C1}^{\prime} (t)} + \frac{h_{C2}^{\prime\prime} (t)}{h_{C2}^{\prime} (t)} = 
2 \left( \frac{h_{C1}' (t)  - h_{C2}' (t)}{h_{C1} (t) - h_{C2} (t)} \right)
\eeq
which integrates to the expected
\beq
h_{C1}' (t) h_{C2}' (t) = -\mathcal{J}^2 (h_{C1} (t) - h_{C2} (t))^2
\eeq
So the final equations for the thermal equilibrium state are
\begin{align}
\frac{h_{C1}^{\prime\prime} (t)}{h_{C1}^{\prime} (t)} &=   \frac{2 h_{C1}^{\prime} (t_1)}{h_{C1} (t) - h_{C2} (t)} 
-\frac{h_{C2}^{\ast\prime} (t)}{h_{C1}^\ast (-\infty) - h_{C2}^\ast (t)} + \frac{h_{C2}^{\prime} (t)}{h_{C1} (-\infty) - h_{C2} (t)} \label{etherm1} \\
& - \frac{h_{C2}^{\prime} (t)}{h_{C1} (t) - h_{C2} (t)} + \frac{ h_{C2}^{\ast\prime} (t)}{h_{C1}^\ast (t) - h_{C2}^\ast (t)} 
\nn
h_{C1}' (t) h_{C2}' (t) &= -\mathcal{J}^2 (h_{C1} (t) - h_{C2} (t))^2 \,. \label{etherm2}
\end{align}
Unlike Eqs.~(\ref{efinal1},\ref{efinal2}), Eqs.~(\ref{etherm1},\ref{etherm2}) have to be integrated from $t=-\infty$.
One solution of Eqs.~(\ref{etherm1},\ref{etherm2}) is
\beq
h_{C1} (t) = h_{C1} (-\infty) + A e^{i \theta} e^{\sigma t} \quad \, \quad  h_{C2} (t) = h_{C1} (-\infty) + A e^{-i \theta} e^{\sigma t} 
\label{Cexact}
\eeq
with
\beq
\sigma = 2 \mathcal{J} \sin (\theta) \,.
\eeq
Note that the $g_C (t_1,t_2)$ obtained from this 
solution agrees with Eq.~(\ref{sp2}) at $\mathcal{J}_2 = 0$ and with 
Eq.~(\ref{sp3}) at $\mathcal{J}_{1/2}=0$.

\bibliography{sykqq.bib}

\end{document}